\documentclass[a4paper,fleqn,useAMS,usenatbib]{mnras}

\pdfoutput=1

\usepackage{natbib}
\usepackage{mathptmx}

\usepackage[T1]{fontenc}
\usepackage{ae,aecompl}

\usepackage[dvips]{graphicx}
\usepackage{amsmath}
\usepackage{amsfonts}
\usepackage{amssymb}
\usepackage{comment}
\usepackage{deluxetable}
\makeatletter
\def\@to{to}
\makeatother
\usepackage{bm} 
\usepackage[dvipsnames]{xcolor}
\usepackage[]{hyperref}
\hypersetup{
	colorlinks=true,	
	urlcolor=MidnightBlue,		
	hypertexnames=true,
	plainpages=false,
	naturalnames=true,
	pdftitle={Binary parameters from astrometric and spectroscopic errors},    
	pdfauthor={Shion Andrew},     
	linkcolor=WildStrawberry,          
	citecolor=Periwinkle,        
}
\def\code#1{\texttt{#1}}
 
 \newcommand\changetwo{}
 \newcommand\changemathtwo{}
\usepackage{amsmath}
\usepackage{amsfonts}
\usepackage{amssymb}
\usepackage{wasysym}
\usepackage[percent]{overpic}
\usepackage{subfig}
\usepackage{csvsimple}
\usepackage{lscape}
\usepackage[shortlabels]{enumitem}
\def\code#1{\texttt{#1}}

\makeatletter
\def\env@matrix{\hskip -\arraycolsep 
  \let\@ifnextchar\new@ifnextchar
  \array{*{\c@MaxMatrixCols}c}}
\makeatother

\DeclareMathAlphabet{\pazocal}{OMS}{zplm}{m}{n}
\newcommand{\unif}{\pazocal{U}}

\newcommand{\rvmethodused}{6,254,790\ }
\newcommand{\fullrvssample}{33,812,183\ }
\newcommand{\bronzelist}{4,641\ }
\newcommand{\silverlist}{2,620\ }
\newcommand{\goldlist}{45\ }
\newcommand{\goldlistinitial}{50\ }

\title[Binary parameters from astrometric and spectroscopic errors]{Binary parameters from astrometric and spectroscopic errors--candidate hierarchical triples and massive dark companions in \textit{Gaia} DR3}
\author[S. Andrew et al.]{Shion Andrew$^{1}$\thanks{E-mail:
\href{mailto:sa2051@cam.ac.uk}{sa2051@cam.ac.uk}}, Zephyr Penoyre$^{1}$, Vasily Belokurov$^{1,2}$, N. Wyn Evans$^{1}$, Semyeong Oh$^{1}$ \\
$^{1}$
Institute of Astronomy, University of Cambridge, Madingley Road, Cambridge, CB3 0HA, United Kingdom\\
$^2$Center for Computational Astrophysics, Flatiron Institute, 162 5th Avenue, New York, NY 10010, USA}
\date{Accepted by MNRAS}

\pubyear{2022}

\begin{document}
\label{firstpage}
\pagerange{\pageref{firstpage}--\pageref{lastpage}}
\maketitle

\begin{abstract}
We show how astrometric and spectroscopic errors introduced by an unresolved binary system can be combined to give estimates of the binary period and mass ratio. This can be performed analytically if we assume we see one or more full orbits over our observational baseline, or numerically for all other cases. We apply this method to \textit{Gaia} DR3 data, combining the most recent astrometric and spectroscopic data. We compare inferred periods and mass ratios calculated using our method with orbital parameters measured for non-single stars in \textit{Gaia} DR3 and find good agreement. Finally, we use this method to search the subset of the \textit{Gaia} DR3 RVS dataset with \code{rv\_method\_used=1} for compact object candidates. We select sources with significant astrometric and spectroscopic errors ($\textit{RUWE}_{ast}>1.25$ and $\textit{RUWE}_{spec}>2$), large inferred mass ratios, and large inferred companion masses ($q>1$ and $m_2>3 M_\odot$) giving a catalogue of \bronzelist candidate hierarchical triples and Main Sequence+Compact Object pairs. We apply more stringent cuts, and impose low levels of photometric variability to remove likely triples ($\textit{RUWE}_{phot}<2$), producing a gold sample of \goldlist candidates. 
\end{abstract}


\section{Introduction}

The \textit{Gaia} survey \citep{Gaia16,Gaia18, Gaia21, Gaia_DR3_Contents} has and will continue to provide astrometric \citep{Lindegren21, Lindegren22}, spectroscopic \citep{Katz22} and photometric \citep{Riello21} measurements for a population of stars orders of magnitude larger than any that existed before. 

The starting assumption for each source is that it is a single star, or at least behaves as such. However, we expect around half of all stars to be in binaries, especially for more massive brighter systems \citep{Offner22}. Long period ($\gtrsim 10$ years) binaries will barely change over the time baseline of the current \textit{Gaia} data releases, though some sufficiently wide binaries may be resolvable as two separate but bound point sources (see for example the catalogue of \citealt{ElBadry21}).

At shorter periods, however, the binary cannot be resolved and can cause extra motion, both in the plane of the sky (detectable via astrometry) and along the line of sight (encoded in spectroscopic measurements). Very short period systems may be close enough to tidally distort the sources of light \citep{Morris1985} and cause a measurable excess of photometric error, although other forms of variability are expected to be common in the dataset as well \citep[e.g.][]{Eyer2019}.

In \citet{Penoyre20}, \citet{Belokurov20}, \citet{Penoyre22a} and \citet{Penoyre22b} we have explored in detail the astrometric contribution of binaries. In this paper we extend a similar line of reasoning, analytically, numerically and observationally, to the spectroscopic and photometric contributions. In particular, we show that for systems which have both significant astrometric and spectroscopic excess--if we assume that these are both caused by the same binary--we can infer the period, masses and semi-major axis of the system.

One of the most exciting uses for this is the selection of candidate Main Sequence (MS) Compact Object (CO) binaries, where the companion is dark (typically, a neutron star or a black hole) but very massive. As shown by others \citep[e.g.][]{Breivik2017,Mashian2017,Yamaguchi_2018,Andrews19,Shahaf19,Chawla21}, \textit{Gaia} is the best available instrument for detecting these systems within our Galaxy. 

Binary-induced astrometric perturbations depend linearly on the semi-major axis of the orbit, and hence become vanishingly small for low period systems (months or less). Thus, the MS+CO systems detectable via astrometric deviations are not the very close binaries which may become interesting gravitational wave sources on human timescales \citep[e.g.][]{Peters1963,Hurley2002}. Nevertheless, they do encode information about potential progenitors to and the stellar evolution of MS+CO binaries in slightly larger orbits \citep{Belczynski2002}.

Any nearby MS+CO systems (with periods ranging from months to years) will likely produce a large and easily detectable excess astrometric and spectroscopic error. However, MS+MS binaries can also induce similar excess errors. Given that these are expected to be many thousand times more common \citep[][]{Breivik2017,Mashian2017}, the difficulty becomes choosing selection criteria which have a high level of specificity. 

In this work, we will show that combining astrometric and spectroscopic measurements can give this high level of specificity. Furthermore, this method can tell us about some of the most physically relevant properties of the system. In Section \ref{method} we outline the analytical framework for translating between binary properties and their associated error, including the dependence on (generally unknown) viewing angles and eccentricity. In Section \ref{sec:simulated_systems}, we apply this to simulated systems, inferring mass ratios and periods based on synthetic \textit{Gaia}-like observations. We do this in part by constructing statistics for spectroscopy and photometry analogous to the astrometric (renormalised) unit weight error\footnote{\textit{RUWE} is the \textit{Gaia} equivalent of the unit weight error (\textit{UWE}), normalised by the corresponding error which in \textit{Gaia} depends on colour and apparent magnitude of the source. We will use the name $\textit{RUWE}_{ast}$ for the renormalised astrometric unit weight error and introduce $\textit{RUWE}_{spec}$ and $\textit{RUWE}_{phot}$ as well.}, both for simulated and observational data in Appendix \ref{ap:gaia_error}. In Section \ref{sec:comparison} we apply our method to \textit{Gaia} data and compare our inferences to known binaries from the APOGEE catalogue \citep{Price-Whelan20}, for which the period and mass ratio have already been measured. Finally in Section \ref{sec:data} we apply our analysis to sources in the \textit{Gaia} Radial Velocity Spectrometer (RVS) sample which have \code{rv\_method\_used=1} and specifically select sources with high inferred mass ratio and companion mass, producing an initial catalogue of \bronzelist candidate hierarchical triples and MS+CO systems. We then apply more stringent cuts on this sample, motivated by estimates of the error on our inferred parameters, and remove samples with a large degree of photometric error to give a gold sample of \goldlist candidate MS+CO systems.  

\begin{figure}
\includegraphics[width=0.49\textwidth]{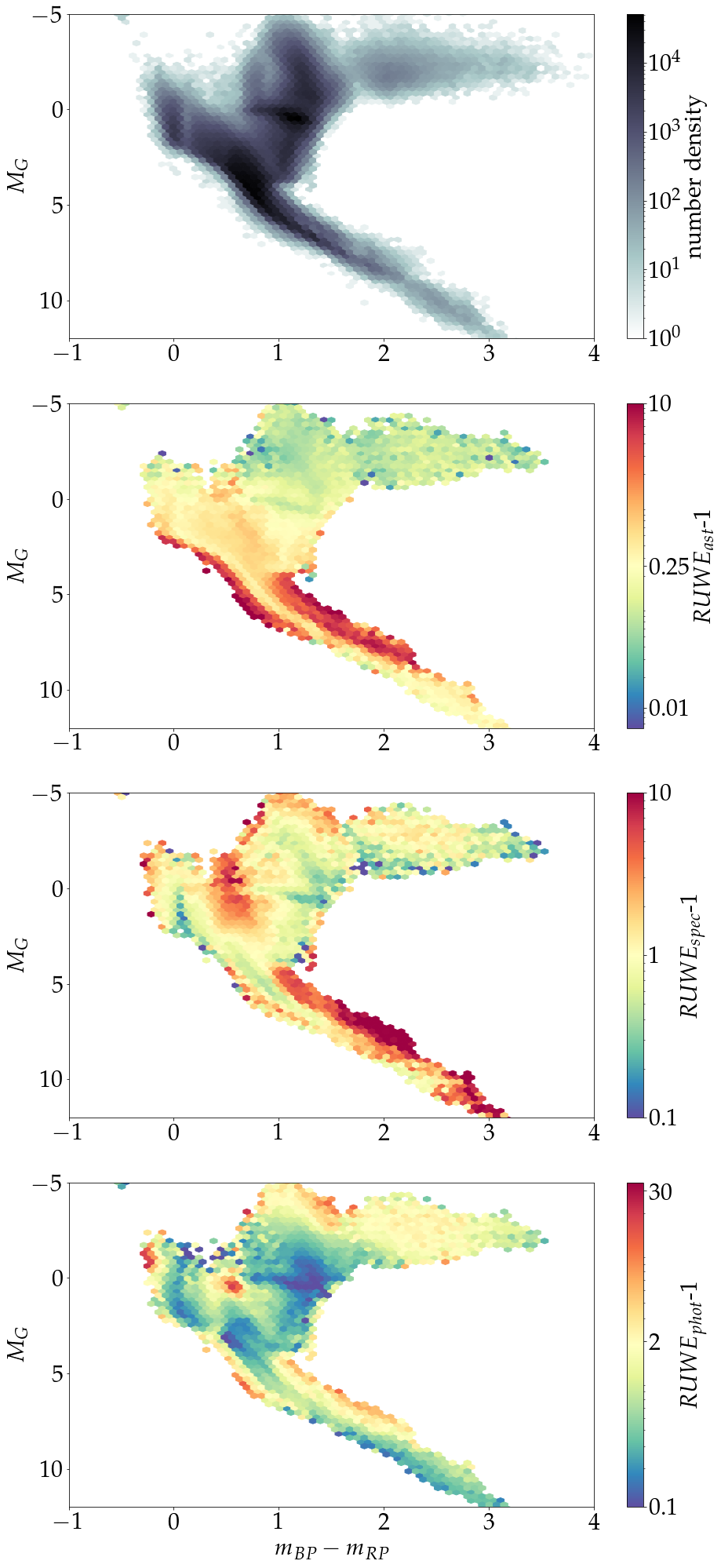}
\caption{Hertzsprung-Russell (HR) diagrams for sources in the \textit{Gaia} RVS sample. We correct for $G$-band extinction and $BP-RP$ reddening using the \code{ag\_gspphot} and \code{ebpminrp\_gspphot} columns provided in the \code{gaia\_source} table. The top panel shows the distribution across colour-magnitude space. Each other panel shows the mean excess \textit{RUWE} (which should have a modal value near to unity for single stars). From top to bottom, we show the astrometric, spectroscopic and photometric contributions. Regions with high \textit{RUWE} are dominated by sources with unmodelled extra contributions of noise, most ubiquitously caused by binary (and higher multiple) systems. Each of the colourmaps is normalised such that the critical value of \textit{RUWE}, above which a source is deemed incompatible with a single star solution, is centred.
}
\label{fig:HRD_RV}
\end{figure}

\section{Method}
\label{method}

To give an initial context to this discussion we start by showing Figure \ref{fig:HRD_RV}, which shows the excess errors in astrometric, radial velocity, and photometric measurements, as parameterised by $\textit{RUWE}_{ast}$ (which is given in \textit{Gaia} DR3 as \code{ruwe}) as well as our constructed $\textit{RUWE}_{spec}$ and $\textit{RUWE}_{phot}$ (as detailed in Appendix \ref{ap:gaia_error}) across the Hertzsprung Russell diagram (HRD). Each of these is a measure of an excess of error compared to a fit performed assuming they are a single non-variable star. 

We will leave a detailed discussion of each of these statistics for later in the paper, but we can already see that in the HRD many of the same regions contain high astrometric, spectroscopic, and photometric excess errors. Most clearly, sitting above the spine of the Main Sequence (MS), we see the MS multiples which have excess astrometric and spectroscopic noise as well as photometric variability detectable by \textit{Gaia}. Other regions, like that occupied by the bright, variable RR Lyrae systems (at $M_G\approx0$ and $BP-RP\approx0.7$) show up clearly in $\textit{RUWE}_{phot}$ without a correspondingly high $\textit{RUWE}_{ast}$ or $\textit{RUWE}_{spec}$ (they are all variable but not, in general, binaries).

Thus, the combination of these statistics can give multiple insights into the properties of a particular star. We will focus particularly on the combination of astrometric and spectroscopic errors (and use photometric variation later only to prune our sample of candidates) and thus start by deriving analytic approximations for the expected contribution of unresolved binary companions.

We note that in this paper, we use the astrometric \textit{RUWE} reported by \textit{Gaia} to extract astrometric errors induced by unresolved binary motion. Although other quality parameters measured by $Gaia$, such as \code{astrometric\_excess\_noise} (AEN) and \code{astrometric\_gof\_al}, could in principal be used as alternative metrics to \code{ruwe}, they are generally well-correlated and quantify the same thing (e.g. see Figure A.2 of \cite{Belokurov20}), providing no distinct advantages over \code{ruwe}. \cite{LindegrenRUWE} also designed the astrometric \textit{RUWE} parameter in order to provide a better goodness-of-fit statistic over these other quality indicators. For instance, AEN contains magnitude/colour-dependent systematic errors (see \cite{Lindegren21}), and \cite{Belokurov20} showed that many sources with high astrometric \textit{RUWE} do not have significant AEN. 

We also note that higher multiples may have even larger signals than binaries. Systems of three or more bodies are quasi-stable, and any long-lived multiple is likely hierarchically arranged \citep{Tokovinin21}. This means that for systems of increasing multiplicity, it is increasingly likely that one component of the multiple has an orbit within any period window. This may be why we see significant photometric variability further above the MS in Figure \ref{fig:HRD_RV} compared to astrometric measurements -- the tight binary orbits needed to produce significant tidal distortions may only become ubiquitous in triples or even higher multiples. These systems, containing more stars, are generally brighter and sit further above the MS (without markedly changing their colour). We will only directly consider binaries in this paper, though we will note when higher multiples may be particularly relevant; they should be considered a major (and interesting) contaminant throughout.

\subsection{Time averaged deviations}

Let us imagine we observe a system over a time baseline $B$\footnote{For example, for the \textit{Gaia} survey's third data release $B_{DR3}=\frac{34}{12}$ years}. If the system is a binary, with period $P$, we observe
\begin{equation}
N_{orb}=\frac{P}{B}
\end{equation}
orbits. If $N_{orb} \gtrsim 1$, we can approximate the behaviour of that system by integrating over one complete orbit, under the assumption of many similarly spaced observations. We will also address the $N_{orb}<1$ case more approximately below and discuss the effects of sparser measurements. This will give a relatively simple analytic form for the expected deviations introduced by a binary -- including both the dimensional scaling with the physical parameters of the system, and a geometric factor (often of order unity) relating to the specific angle and phase at which the system is viewed.

The relevant parameters of the binary, which may effect any observed quantity, are the period $P$, semi-major axis $a$, and  eccentricity, $e$. We designate the brighter star as the primary (as it is the one we observe) with luminosity ratio $l\ (\leq 1)$ and mass ratio $q=\frac{m_2}{m_1}$ where $m_2$ is the mass of the companion and $m_1$ the mass of the primary. Through Kepler's third law any two of the three parameters $P$, $a$, and $m_1(1+q)$ (the total mass) are sufficient to set the value of the third via
\begin{equation}
\label{eqn:kepler}
\left(\frac{P}{2\pi}\right)^2 = \frac{a^3}{Gm_1(1+q)}.
\end{equation}
Similarly, $m_2$ is specified entirely by $m_1$ and $q$, and need not enter in to our calculations (though it may be interesting to examine as an end-product).

Measured quantities also depend on observational parameters that do not alter the fundamental physics of the system: these are the viewing angles $(\theta_v,\phi_v,\omega_v)$, parallax, $\varpi$, time of periapse $t_p$, and the relevant observational errors, $\sigma_{spec}$ and $\sigma_{ast}$ for spectroscopic and astrometric measurements respectively. Following the naming convention given in \cite{Penoyre20}, $\theta_v$ is defined to be the inclination angle between the orbital plane and the line-of-sight vector, and $\phi_v$ is defined to be the angle between the axis of periapse and the line-of-sight vector projected onto the orbital plane. One of the viewing angles, $\omega_v$, specifies the orientation of the binary relative to the measurement co-ordinate system, and thus will not change any observable results unless the measurements are anisotropic (another assumption that can be violated in practice by \textit{Gaia}'s  scanning law, but which we will assume is satisfied for now).

Thus, there are 5 independent physical parameters of the system ($P,\ m_1,\ q,\ l,\ e$), 4 viewer specific parameters ($\theta_v,\ \phi_v,\ t_p,\ \varpi$) and 2 errors ($\sigma_{ast},\ \sigma_{spec}$) which define the excess error introduced by a binary system and its significance over our measurement sensitivity.

\subsection{Error introduced by a binary}

\begin{figure*}
\centering
\includegraphics[width=0.98\textwidth]{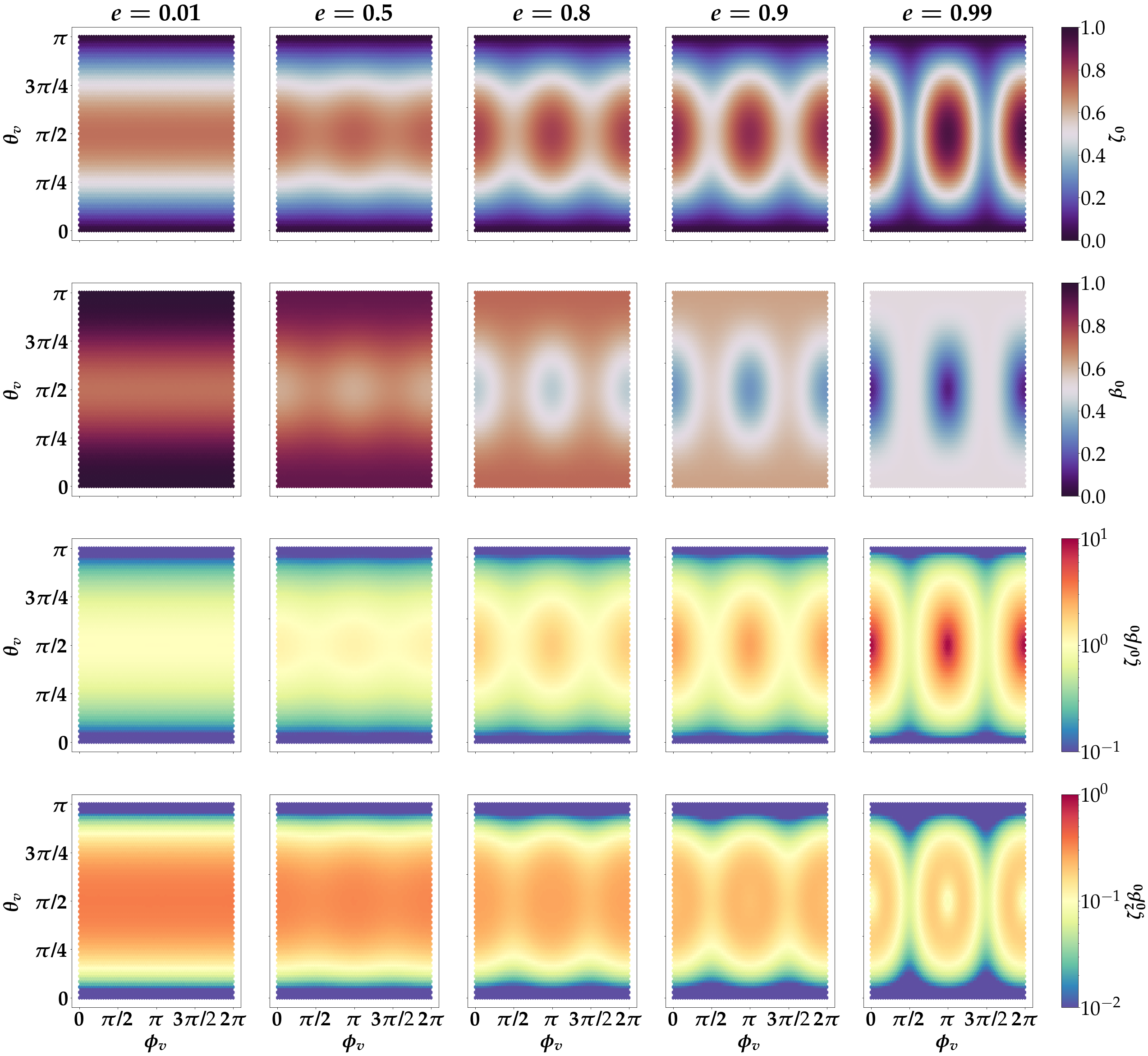}
\caption{The behaviour of the simple forms of the astrometric ($\beta_0$) and spectroscopic ($\zeta_0$) projection terms for different viewing angles and eccentricities. We also show the factor $\frac{\zeta_0}{\beta_0}$ which appears in the formula for period $P$ (Equation \ref{period}), and $\zeta_0^2 \beta_0$ which directly affects the mass ratio $q$ (Equations \ref{qalpha} and \ref{qfullsolution}). $\zeta_0$ tends towards zero when the binary is viewed face-on ($\sin\theta_v=0$) and is maximised at values of $\phi_v$ when most of the binary motion is radial. We see the inverse behaviour with $\beta$, which is maximised when most of the binary motion is planar and minimised when the binary motion is primarily along the line of sight. $\zeta_0/\beta_0$ is in general of order unity, while $\zeta_0^2\beta_0$ can take on very small ($\sim 10^-2$) values.}
\label{theoretical_beta_zeta_grid}
\end{figure*}

\begin{figure}
\centering
\includegraphics[width=0.49\textwidth]{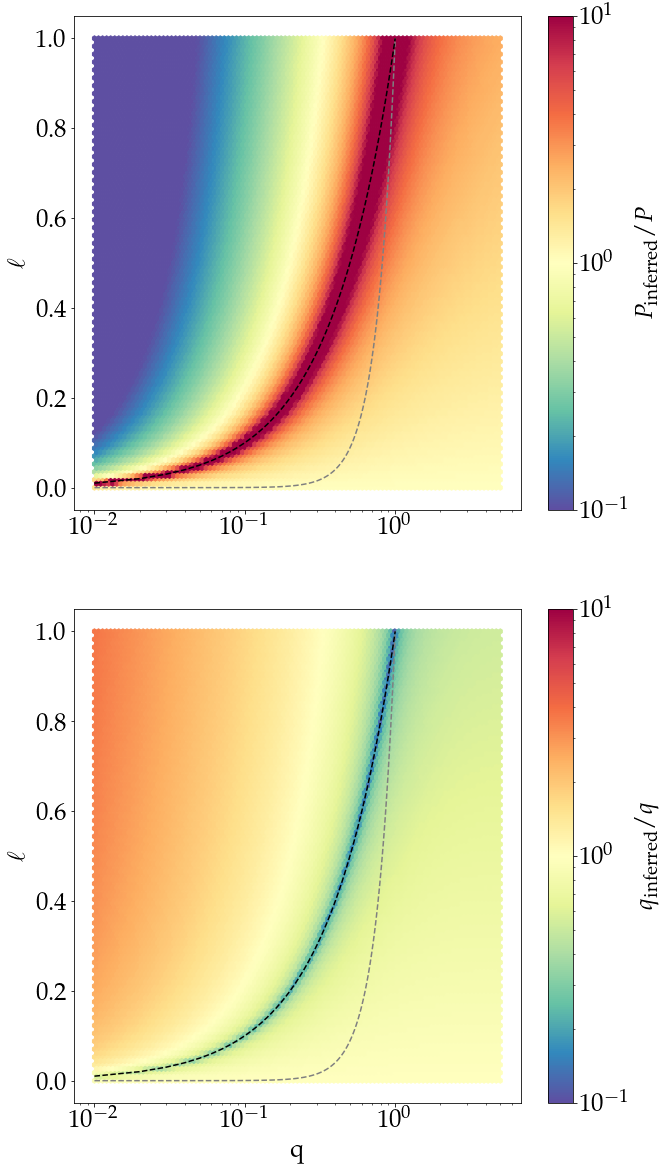}
\caption{Top: Inferred period assuming $l=0$ (Equation \ref{period}) over the true period (Equation \ref{period_general}) in $l-q$ space. A black dotted-line is drawn at $l=q$ and a grey dotted-line is drawn at $l=q^{3.5}$ for MS-MS binaries to guide the eye. Bottom: same as above, except colour-coded by inferred mass ratio assuming $l=0$ (Equation \ref{eqn:q_cubic}) over true mass ratio.}
\label{fig:l_v_q_massratio}
\end{figure}

We will focus on two error terms, $\sigma_{b,vr}$ and $\sigma_{b,\theta}$, corresponding to the error introduced by a binary in spectroscopic and astrometric measurements.

If we imagine a known binary, and assume it is frequently and isotropically scanned, we can derive analytic forms for these assuming $P=B$. As we will show, this result agrees well for all $P<B$ but deviates at longer periods.

\subsubsection{Spectroscopic error}

The standard deviation of measurements of the radial velocity of a binary system, which we derive in Appendix \ref{ap:rvscatter}, follows the form
 \begin{align}
    \label{eqn:sigma_rv}
     \sigma_{b,vr} 
    &=\frac{q}{1+q}\cdot \frac{2 \pi a}{P}\cdot \zeta(P,e,t_p,\theta_v,\phi_v)
 \end{align}
 where $P$ denotes the binary period, $a$ denotes the semi-major axis, and $\zeta$ is a function that depends on the viewing angle, eccentricity, orbital period, and observational baselines. In the case that $P=B$, the analytic expression for $\zeta$, which we will call $\zeta_0$, is 
\begin{equation}
    \label{eqn:zeta0}
    \zeta_0(e,\theta_v,\phi_v) = \frac{1}{e}\sqrt{\kappa_{ss}^2\epsilon(1-\epsilon)-\kappa_{sc}^2\frac{(e^4+e^2(3\epsilon-5)+4(1-\epsilon))}{ (1-\epsilon)^2\epsilon}}
\end{equation}
where $\kappa_{ss}=\sin\theta_v\sin\phi_v$, $\kappa_{sc}\equiv\sin\theta_v\cos\phi_v$, and $\epsilon=\sqrt{1-e^2}$. 

The top row of Figure \ref{theoretical_beta_zeta_grid} shows $\zeta_0$ as a function of viewing angle and eccentricity. It is always small for face-on systems ($\sin\theta_v$ close to zero) and also for eccentric systems where the observed motion is along the line of sight at apoapsis and periapse ($\cos\phi_v$ is small).

\subsubsection{Astrometric error}

As shown in \citet{Penoyre20} and \citet{Penoyre22a}, the expected astrometric error introduced by a binary is
\begin{equation}
\label{eqn:sigma_theta}
\sigma_{b,\theta}=|\Delta_{q,l}| \cdot \varpi \cdot \frac{a}{A} \cdot \beta(P,e,t_p,\theta_v,\phi_v)
\end{equation}
where $A$ is one astronomical unit and $\beta$ is a factor of order unity for $P\lesssim B$. $\beta$ encodes all dependence on the viewing angle, and
\begin{equation}
\Delta_{q,l}=\frac{q-l}{(1+q)(1+l)}
\end{equation}
is the relative offset between the photocentre and the centre of mass.

In the case that $P=B$, the analytic expression for $\beta$, which we will call $\beta_0$, was shown by \citet{Penoyre22a} to be
\begin{equation}
\label{eqn:beta0}
\beta_0(e,\theta_v,\phi_v)=\sqrt{1-\frac{\sin{\theta_v}^2}{2} - e^2 \frac{3+\sin{\theta_v}^2(\cos{\phi_v}^2-2)}{4}}.
\end{equation}

The second row of Figure \ref{theoretical_beta_zeta_grid} shows $\beta_0$ as a function of viewing angle and eccentricity. As discussed in \citet{Penoyre22a}, the value is significantly greater than 0 for all but very eccentric systems aligned such that the motion is primarily along-the line of sight ($\sin\theta_v$ close to 1, and $\cos\phi_v$ close to zero).

\subsection{Physical parameters from measured errors}
\label{sec:Parameters_from_measured_errors}

We can invert Equations \ref{eqn:kepler}, \ref{eqn:sigma_rv}, and \ref{eqn:sigma_theta} to recover the period of a binary system. This is particularly easy to do in cases where the companion is dark ($l=0$) at which point $|\Delta_{q,l}|=\frac{q}{1+q}$ and the period is
\begin{equation}
\label{period}
P=\frac{2 \pi A}{\sigma_{b,vr}} \frac{\sigma_{b,\theta}}{\varpi} \frac{\zeta}{\beta}.
\end{equation}
The mass ratio can also be expressed, in the form of a cubic, as
\begin{equation}
\label{eqn:q_cubic}
q^3 -\alpha q^2 - 2\alpha q - \alpha = 0
\end{equation}
where
\begin{equation}
\label{qalpha}
\alpha=\frac{A \sigma_{b,vr}^2}{Gm_1}\frac{\sigma_{b,\theta}}{\varpi} \frac{1}{\beta \zeta^2}.
\end{equation}

This is equivalent to the expression given in Equation 4 of \citet{Shahaf19} (where their $\mathcal{A}$ is equal to our $\alpha^\frac{1}{3}$). Thus, under the assumption of negligible $l$, we would be able to perform a similar analysis but without needing to know the parameters of the binary, just the associated errors. However, as we'll detail in Section \ref{sec:triples}, our estimation breaks down for triples and we cannot use their criteria to separate higher multiple systems from compact object companions.

Equation \ref{eqn:q_cubic} has one real root (given that $\alpha>0$ always) which can be found through translating to a depressed cubic and using Cardano's formula to give
\begin{equation}
\label{qfullsolution}
q=\frac{\alpha}{3}+\left(\frac{\lambda}{2}\right)^{\frac{1}{3}}\left[
\left(1+\sqrt{1+\frac{4 \mu^3}{27 \lambda^2}} \right)^{\frac{1}{3}}
+\left(1-\sqrt{1+\frac{4 \mu^3}{27 \lambda^2}} \right)^{\frac{1}{3}}
\right]
\end{equation}
where
\begin{equation}
\mu(\alpha)=-\frac{6+\alpha}{3}\alpha
\end{equation}
and
\begin{equation}
\lambda(\alpha)=\frac{27+18\alpha+2\alpha^2}{27}\alpha.
\end{equation}
We note \changetwo{that $\changemathtwo{q\ge \alpha}$ always}, and for extreme values of $\alpha$, we get the following asymptotic equivalences:
\begin{align}
&q \sim  \alpha^{\frac{1}{3}} \ \ \ (\rm{as\ } \alpha\rightarrow0)
\\
&q \sim \alpha \ \ \ \ (\rm{as\ } \alpha\rightarrow\infty).
\end{align}
In general calculating $\alpha$, and thus $q$, requires a known value of the mass of the primary $m_1$. However, it's interesting to note that as $\alpha \propto m_1^{-1}$ for large $\alpha$, the mass of the secondary $m_2=q m_1$ becomes independent of the primary mass. 

The third row of Figure \ref{theoretical_beta_zeta_grid} shows $\zeta_0/\beta_0$ (the expected dependence on the viewing angle factors that appear in Equation \ref{period} for calculating the period). The fourth row shows $\zeta_0^2 \beta_0$ (which is used to estimate the mass ratio). The former can be significantly larger than the latter for eccentric orbits when $\beta_0$ can take very small values. Both go to zero for face-on orbits (when there is effectively no radial motion and $\sigma_{b,vr} \rightarrow 0$) and $\zeta_0^2 \beta_0$ has a complex 'doughnut'-like structure at high eccentricities.
 
In the more general case that $l \neq 0$, instead we have
 \begin{equation}
 \label{period_general}
 P=\frac{2\pi A}{\sigma_{b,vr}}\frac{\sigma_{b,\theta}}{\varpi}\frac{\zeta}{\beta}\frac{q(1+l)}{|q-l|}
 \end{equation}
 and for the mass ratio
 \begin{equation}
  \label{q_general}
q^2|q-l|-\alpha(1+q)^2(l+1)=0.
\end{equation}
 In Figure \ref{fig:l_v_q_massratio}, we use these more general Equations to find the regions in $l-q$ space in which Equations \ref{period} and \ref{eqn:q_cubic} underestimate and overestimate the true periods and mass ratios. For MS-MS binaries ($l \approx q^{3.5}$), we expect periods on average to be overestimated and mass ratios to be underestimated. In contrast, for binaries with large light ratios relative to their mass ratios (i.e. $l\gg q$), we expect periods to be underestimated and mass ratios to be overestimated. 

 When we apply our method to real data in Section \ref{sec:data} we work under the assumption of $l=0$. We flag candidate MS-CO binaries based on the criteria that inferred mass ratios and secondary masses are large: for MS-MS binaries (which we expect to be the most ubiquitous type of binary in our initial data set), our approximation should result in mass ratios that are underestimated (therefore these systems will be removed with a cut on mass ratio and secondary mass). 
 
\subsection{Triples and false positives}
\label{sec:triples}
The method described in Section \ref{sec:Parameters_from_measured_errors} breaks down in the case of an unresolved triple or higher-order multiple where the astrometric and spectroscopic errors are induced by different orbital periods. 

For instance, in the case of a hierarchical triple, if the inner orbital period is very short (e.g. $P_{b_1} \lesssim 1$ day) we would expect the astrometric error due to the inner orbit $\sigma_{b_1,\theta}$ to be negligible (recall from Equation \ref{eqn:sigma_theta} that $\sigma_{b,\theta} \propto a \propto P^\frac{2}{3}$) and the spectroscopic error due to the inner orbit $\sigma_{b_1,vr}$ to be high (from Equation \ref{eqn:sigma_rv}, $\sigma_{b,vr}\propto P^{-1}$). However, if the outer orbital period ($P_{b_2}$) is on the order of a few years, we expect the astrometric error due to the outer orbit, $\sigma_{b_2,\theta}$, to be large and the spectroscopic error due to the outer orbit, $\sigma_{b_2,vr}$, to be small. Generally, we cannot distinguish whether the astrometric and spectroscopic errors are induced by the same or different orbits. 

Therefore, if both the observed spectroscopic error and the astrometric error are large as a result of being induced by different orbits such that we measure 
\begin{align*}
    \sigma_{vr}=\sigma_{b_1,vr} \\
    \sigma_{\theta}=\sigma_{b_2,\theta},
\end{align*}
then our inferred period 
\begin{equation}
    P_{\rm{inferred}}\propto \frac{\sigma_{b_1,vr}}{\sigma_{b_2,\theta}}
\end{equation}
would be an intermediate value ($P_{b_1}<P_{\rm{inferred}}<P_{b_2}$).

Furthermore, we would also observe a larger $\alpha$ ($\propto \sigma_{vr}^2 \sigma_{\theta}$, Equation \ref{qalpha}) and therefore infer a larger mass ratio than the true value for either orbit. Therefore, while selecting for sources with large inferred mass ratios can be used to obtain candidate MS+CO binaries, it can also be used to identify unresolved hierarchical triples. This is discussed further with respect to \textit{Gaia} data in Section \ref{sec:data}.

Finally, we also note that there are causes unrelated to orbital motion that can induce high astrometric and spectroscopic errors. These include those that are astrophysical by nature, such as Variability Induced Movers \citep{Wielen96}, and those that are non-astrophysical by nature, such as over-crowding on the sky. While quality cuts can minimise such contaminants, some un-modelled instrumental noise from the telescope resulting in spurious measurements is inevitable. For instance, it is also possible for the re-normalisation factors ($\sigma_{ast},\sigma_{spec}$ in Equations \ref{eqn:RUWE} and \ref{eqn:RUWE_Spec}) to be  underestimated for some sources, resulting in inflated \textit{RUWE} values. However, we generally expect $\textit{RUWE}_{ast}$ and $\textit{RUWE}_{spec}$ to have different contaminants, making it significantly less likely for both \textit{RUWE} values to be inflated for an individual source.

\begin{deluxetable}{c c c}
\tabletypesize{\footnotesize}
\tablecolumns{8}
\tablewidth{\linewidth}
\tablecaption{ Simulated Parameter Distribution \label{table:simulated_parameter_distribution}}
\tablehead{\colhead{Parameter} & \colhead{Description} & \colhead{Distribution}} 
\startdata
$\alpha$ (deg) & right ascension &$\unif(0,360)$ \\
$\delta$ (deg) & declination & $\frac{180}{\pi}\sin^{-1}\unif(-1,1)$ \\
$\varpi$ (mas) & parallax & 0.5$\cdot\unif(0,1)^{-1/3}$  \\
$\mu_{\alpha *}$ (mas/yr) & proper motion along ra &  $\varpi \cdot \mathcal{N}$(0,6.67)\\
$\mu_\delta$ (mas/yr) & proper motion along dec &$\varpi \cdot \mathcal{N}$(0,6.67)\\
$\phi_v$ (rad) &azimuthal viewing angle & $\unif(0,2\pi)$ \\
$\theta_v$ (rad) &polar viewing angle & $\cos^{-1}\unif(-1,1)$ \\
$\omega_v$ (rad) &planar projection angle & $\unif(0,2\pi)$\\
$e$ &eccentricity  & $\unif(0,1)$\\
$T_0$ (yr) & time of periapse & $P \cdot \unif(0,1)$ \\
$P$ (yrs) & period & $10^{\pazocal{U}(-2.3,2.0)}$ \\
$m_1$ ($M_\odot$) & mass of primary & 1$M_\odot$\\
$q$ & mass ratio  & $10^{\pazocal{U}(-3\textbf{}.2,2.2)}$
\enddata
\vspace{-0.3cm}
\tablecomments{Distribution of parameters for simulated  MS+CO binary systems. The light ratio $l$ is set to zero for all binaries, and the mass of the secondary $m_2$ calculated from the mass of the primary $m_1$ and the mass ratio $q=m_2/m_1$. The semi-major axis is calculated using Kepler's 3rd law along with the period and total mass of each binary.}
\end{deluxetable}

\section{Inferring mass ratios and periods with simulated systems}
\label{sec:simulated_systems}

\begin{figure*}
\centering
\includegraphics[width=0.98\textwidth]{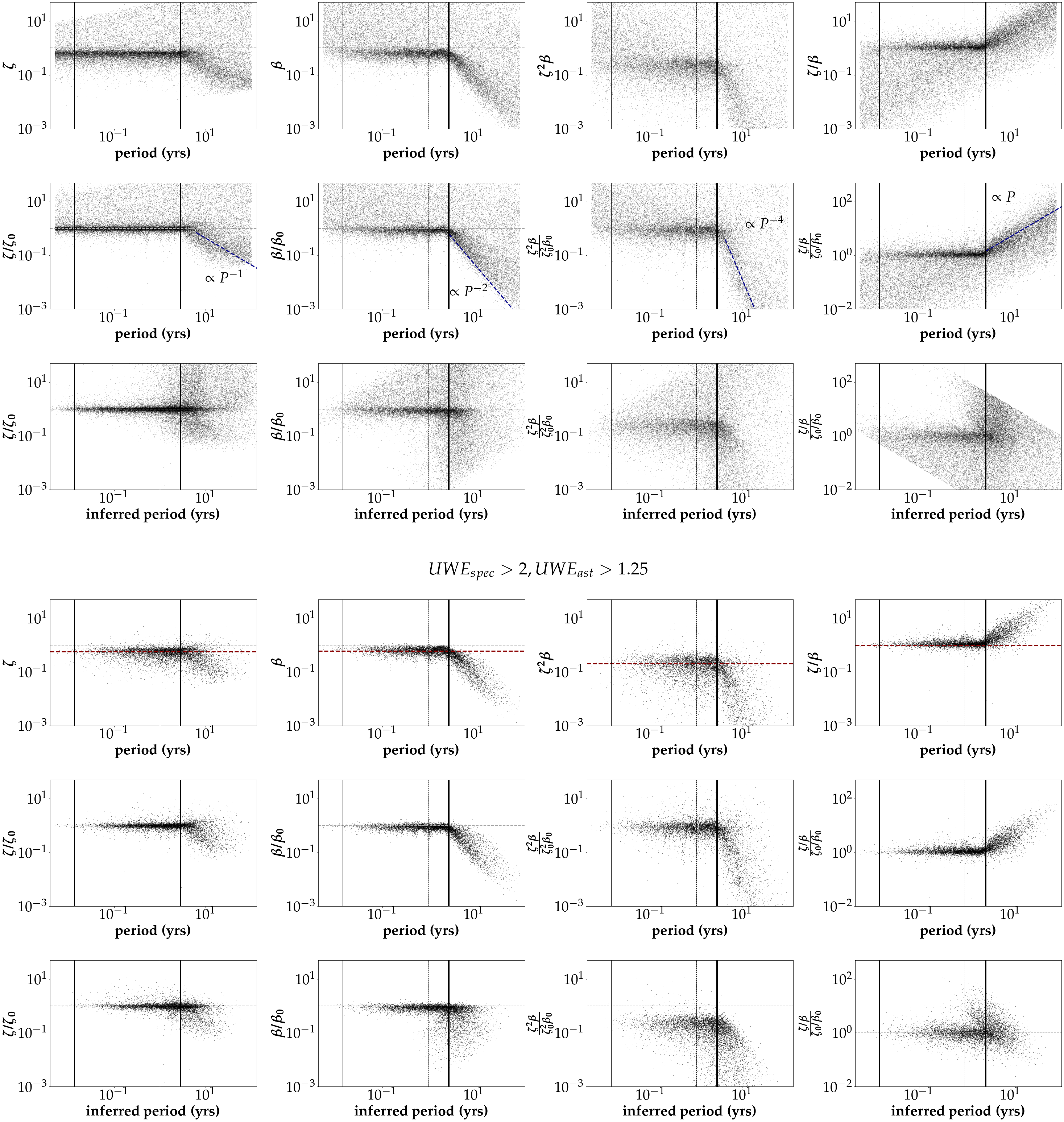}
\caption{Top panels: True values of $\beta$, $\zeta$ as a function of period for simulated systems, calculated by inverting Equations \ref{eqn:sigma_rv} and \ref{eqn:sigma_theta}. Vertical solid lines are plotted at 5 days, 1 year, and 34 months (observational baseline for radial velocity measurements). Horizontal lines are plotted at 1. Second and third row of panels: True values of $\beta$, $\zeta$ normalised by $\beta_0$ and $\zeta_0$ (given by Equations \ref{eqn:beta0} and \ref{eqn:zeta0}) as a function of period (second row of panels) and inferred period (third row of panels) for simulated systems. Dashed blue lines are plotted in 2nd row of panels to extrapolate the behaviour of $\beta$ and $\zeta$ at periods beyond the observational baseline; $\beta$ decays approximately as $P^{-2}$, and $\zeta$ decays approximately as $P^{-1}$ until the noise floor is reached. The bottom three rows of panels are repeats of the top three rows of panels with cuts of $\textit{RUWE}_{ast}>1.25$ and $\textit{RUWE}_{spec}>2$. Horizontal red lines are drawn at the approximate median values of  $\beta=0.62,\zeta=0.55,\zeta^2\beta=0.2,\zeta/\beta=1$ on the 4th row of panels.}
\label{simulated_beta_zeta_period}
\end{figure*}

\begin{figure}
\centering
\includegraphics[width=0.45\textwidth]{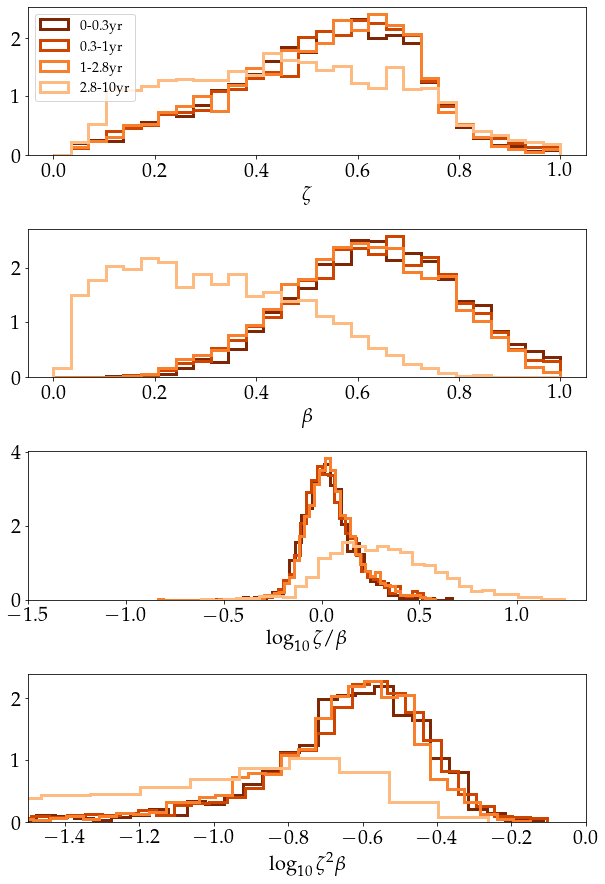}
\caption{Distribution of $\beta, \zeta, \zeta/\beta,\zeta^2\beta$ for simulated sources with $\textit{RUWE}_{ast}>1.25$ and $\textit{RUWE}_{spec}>2$. The distributions are colour-coded by orbital period. We see an excess in low values of $\zeta$ and $\beta$ for long period ($>34$ months) systems that survive the $\textit{RUWE}_{ast}>1.25$ and $\textit{RUWE}_{spec}>2$ cuts. For sources with $P<34$ months, the values of $\zeta$ and $\beta$ (when all eccentricities and viewing angles are equally probable) are well-represented by the same distributions. }
\label{fig:beta_zeta_hist}
\end{figure}

\begin{figure}
\includegraphics[width=0.48\textwidth]{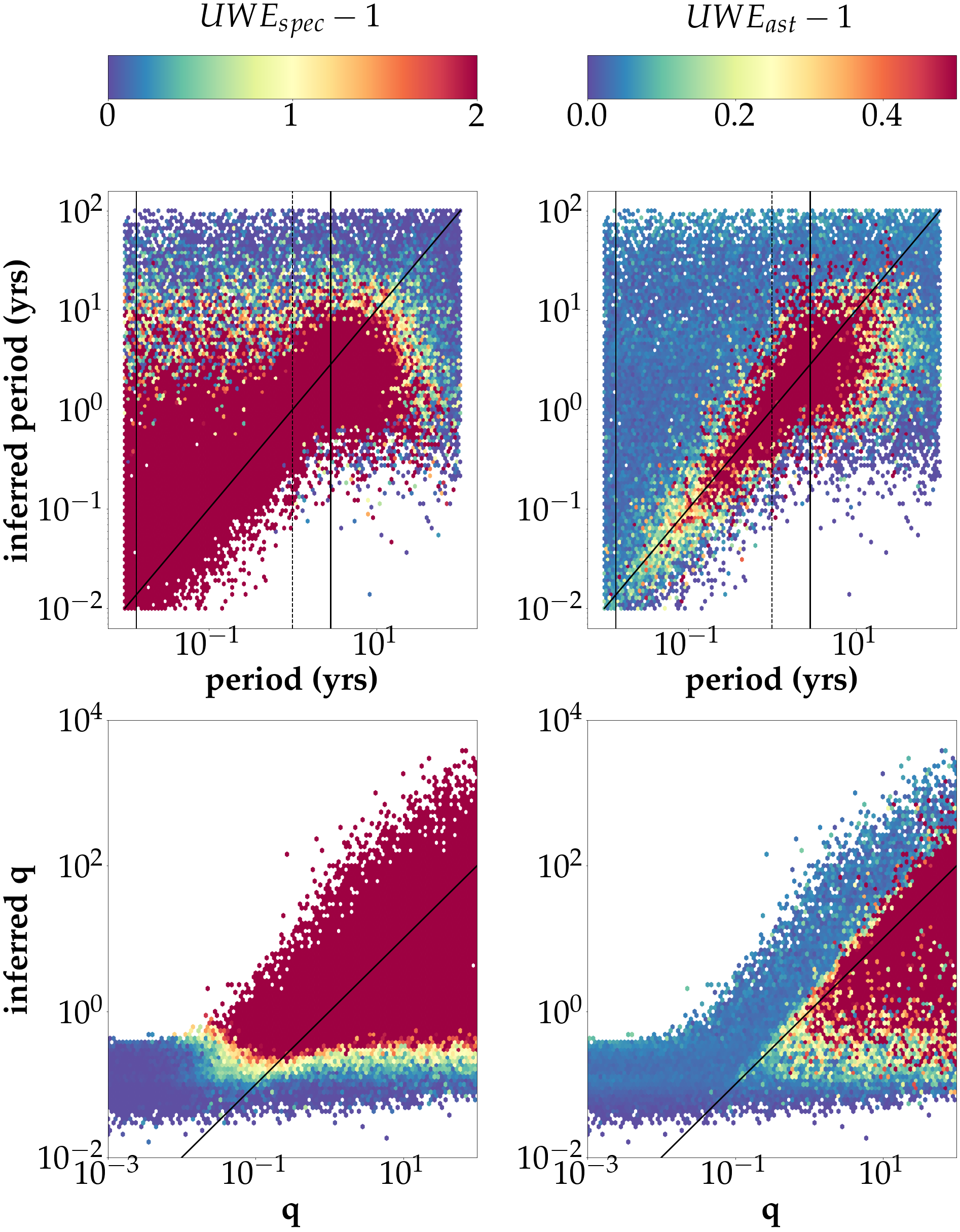}
\caption{Top panels: Inferred period versus true period for simulated data colour coded by $\textit{UWE}_{spec}$ (left) and $\textit{UWE}_{ast}$ (right). Vertical lines are drawn at 5 days, 1 year, and 34 months, and the diagonal corresponds to perfect agreement between the true period and our approximation. Bottom panels: Inferred q versus q for simulated data.}
\label{fig:ruwes_simulation_inferred_q}
\end{figure}

\begin{figure}
\includegraphics[width=0.48\textwidth]{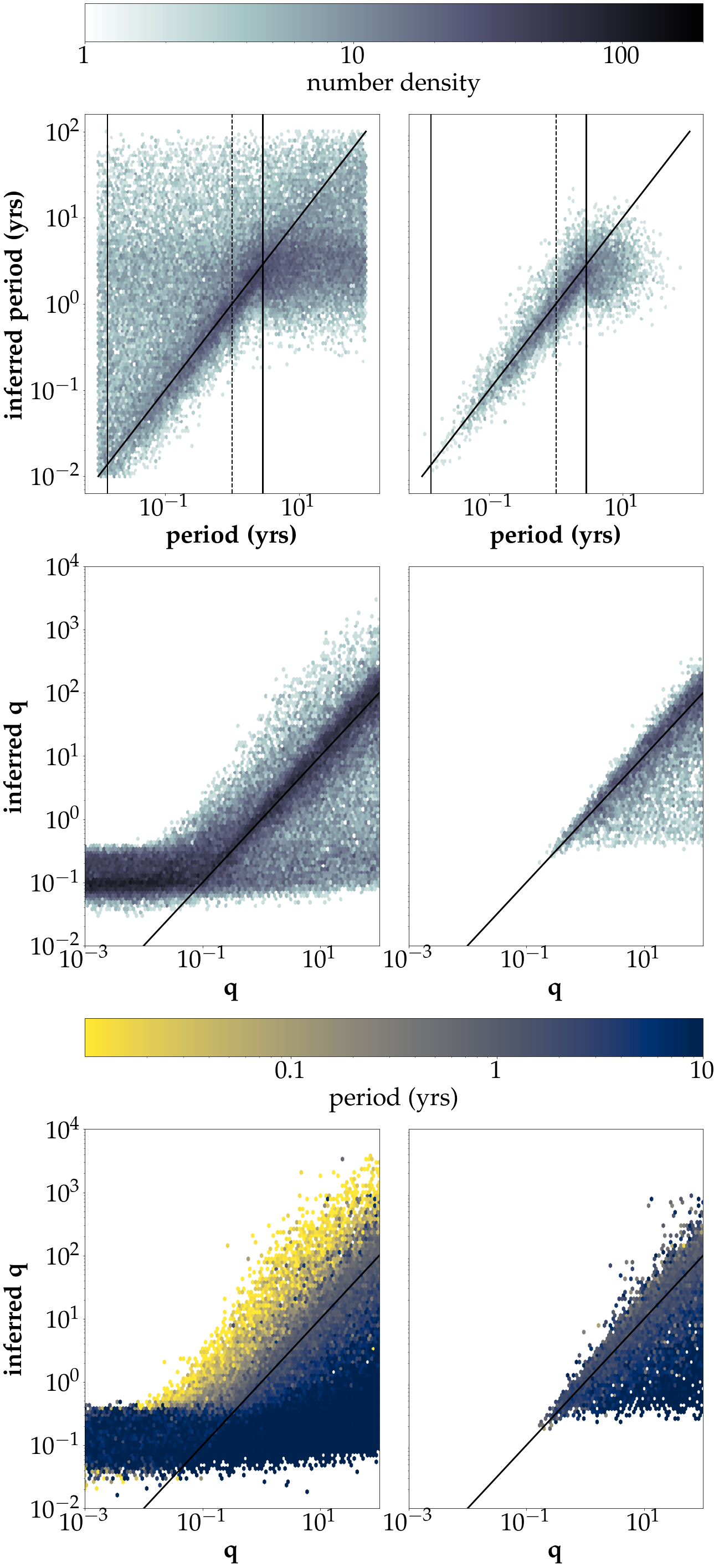}
\caption{Inferred periods and mass ratios versus true values for entire simulated data (left column) and after $\textit{UWE}_{ast}>1.25$ and $\textit{UWE}_{spec}>2$ cuts (right column). Vertical lines are shown at 5 days, 1 year, and 34 months for he top row.}
\label{fig:simulation_inferred_q}
\end{figure}

\begin{figure}
\centering
\includegraphics[width=0.4\textwidth]{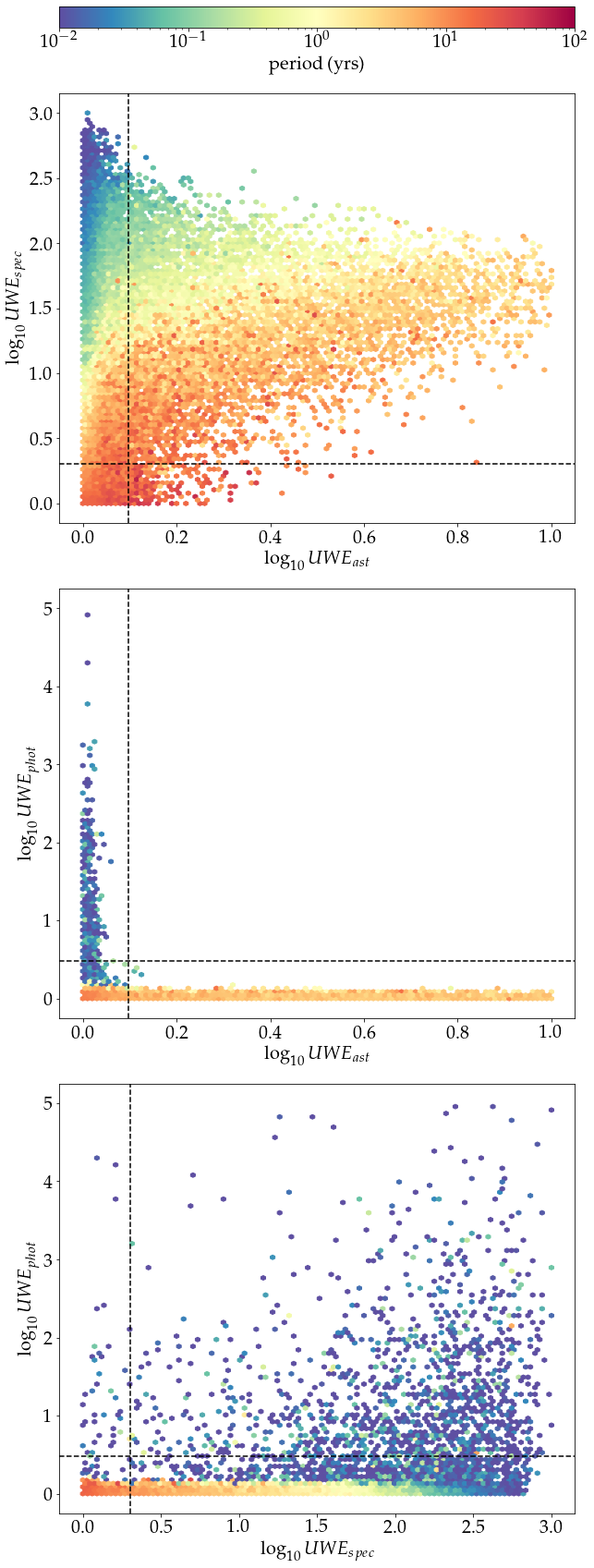}
\caption{Distribution of combinations of $\textit{UWE}_{spec}$, $\textit{UWE}_{ast}$ and $\textit{UWE}_{phot}$ for simulated MS+CO systems colour coded by orbital period. Vertical and horizontal lines show the cuts for significant values for each type of \textit{UWE}. }
\label{fig:RUWE_phot}
\end{figure}

To test this method, we have simulated two million MS+CO systems within 2kpc using \texttt{astromet.py} \footnote{https://github.com/zpenoyre/astromet.py}, a software developed by \cite{Penoyre22a} which emulates \textit{Gaia}’s single-body astrometric fitting pipeline. We also use the \texttt{scanninglaw} package\footnote{https://github.com/gaiaverse/scanninglaw} \citep{Everall_2021,Boubert_2021,Green_2018} that predicts \textit{Gaia} observation times and scanning angles as a function of locations on the sky.

The parameters for our simulated sources are drawn from the distributions given in Table \ref{table:simulated_parameter_distribution}. All stars are uniformly distributed across the sky and out to a distance of 2 kpc. We note that our assumption that observations cover all \changetwo{position angles} in the sky uniformly is unrealistic along the ecliptic due to the presence of the sun, and in some pathological cases can result in excess noise due to a very uneven scanning law--however, these cases are rare and have minimal impact on the results of our simulations.  A normal distribution is used to approximate the observed distribution of proper motions for the bulk of stars in \textit{Gaia} data. All possible viewing angles and orbital eccentricities are assumed to be equally probable. We hold the mass of the primary fixed at $1 M_\odot$ and vary the mass of the secondary by sampling from a log-uniform mass ratio distribution. The log-uniform period distribution was chosen to capture the behaviour of very short ($P\sim 1$ day) and long  ($P>10$ yr) period binaries. While many of our distributions are simple and approximate, the purpose of this section is just to illustrate the effectiveness of inferring mass ratios and periods from measured astrometric parameters alone. A flat along-scan astrometric error $\sigma_{ast}$ of 0.3 mas was adopted based off the expected uncertainties for sources with apparent magnitudes $m_G<16$ shown in Figure A.1 of \cite{Lindegren21}, and radial velocity noise floor of $1.0$ km/s was chosen based off the distribution of $\sigma_{spec}$ for \textit{Gaia} data shown in Figure \ref{fig:sigmaspec_bprp_g}. 

The top panel of Figure \ref{simulated_beta_zeta_period} shows $\zeta$ and $\beta$--calculated inverting Equations \ref{eqn:sigma_rv} and \ref{eqn:sigma_theta}--as a function of period for our simulated systems. All measurements were taken over the baseline of DR3 (34 months). We see that $\beta$ and $\zeta$ are of order unity between $\sim 5$days and 34 months, and drop significantly for sources with periods longer than the observational baseline. 

The second row of Figure \ref{simulated_beta_zeta_period} shows that the values of $\zeta$ normalised by $\zeta_0$ are approximately 1 for sources with periods less than 34 months.  The values of $\beta/\beta_0$ are noticeably less than 1, which follows from the fact that the simple form of astrometric deviations calculated in \citet{Penoyre20} and \citet{Penoyre22a} do not account for the capacity of a binary to bias the single body solution (most notably altering the proper motion), which will in turn reduce the residuals between the observed motion and the fit. Thus, our predicted $\beta$ should be seen as an upper limit.

We also show the distributions of $\zeta^2 \beta$, which is used to calculate inferred mass ratios, and $\zeta/\beta$, which is directly proportional to the inferred period. The values of these parameters, even when normalising by $\zeta_0$ and $\beta_0$, change precipitously for periods beyond the baseline of the survey. One way to understand this is to approximate the behaviour at long periods as a polynomial in $B/P$, for which the low order terms will dominate for $N_{orb} \ll 1$. For the spectroscopic measurements, the constant term is subsumed by the fitted radial velocity and hence the first order term is the lowest order remaining term. For astrometry, this first order is similarly subsumed into proper motion, so the remaining excess is $\propto P^{-2}$. From these scalings, we can also estimate $\zeta^2 \beta \propto P^{-4}$ and $\zeta/\beta \propto P$. This latter result is significant because now both sides of Equation \ref{period} are proportional to $P$ and we can no longer find a reliable estimate for the period. We can clearly see this behaviour in the third row of Figure \ref{simulated_beta_zeta_period} where the inferred periods seem to prefer values just above the baseline of the survey.

At long periods, the noise floor dominates the binary contribution to the radial velocity error, and an uptick in $\zeta$ occurs (since $\zeta \propto P^{1/3}$ when $\sigma_{b,vr}$ and $q$ are constant) for $P>10$ years. Similarly at low periods the astrometric signal becomes noise dominated, a behaviour we start to see below periods of a few days.

\cite{Penoyre20} showed that the astrometric error introduced by a binary is related to the unit weight error as 
\begin{equation}
    \label{eqn:RUWE}
    UWE=\sqrt{\frac{N}{N-5}}\sqrt{1+\frac{1}{2}\bigg(\frac{\sigma_{b,\theta}}{\sigma_{ast}}\bigg)^2}
\end{equation}
where $N$ is the number of observations and $\sigma_{ast}$ is the astrometric error. \cite{Penoyre20} also showed that a cut of $\textit{UWE}_{ast}>1.25$ can be used to identify sources with ``significant" astrometric deviations that are inconsistent with single source astrometry in \textit{Gaia} DR3. We define an analogous \textit{UWE} for $\sigma_{b,vr}$ which we denote $\textit{UWE}_{spec}$, and derive a similar cut of $\textit{UWE}_{spec}>2$ in \ref{ap:rvs_simulated_ast}. The bottom three rows of Figure \ref{simulated_beta_zeta_period} show that, when removing sources with $\textit{UWE}_{ast}<1.25$ and $\textit{UWE}_{spec}<2$, a significant amount of the scatter is removed, indicating that this method is only suitable for sources with significant astrometric and spectroscopic errors.

We can also estimate some fiducial values for $\beta$ and $\zeta$ from this cleaner subsample. We find $\beta_{median}=0.62$ and $\zeta_{median}=0.55$. If we used information about the viewing angle for each system, we could find more precise and motivated values of $\beta$ and $\zeta$. In general, however, these are not known, and we will assume no foreknowledge of the viewing angle throughout the rest of this work. Instead, we randomly draw from the distributions of $\zeta$ and $\beta$ shown in Figure \ref{fig:beta_zeta_hist} for sources with $\textit{UWE}_{ast}>1.25$, $\textit{UWE}_{spec}>2$, and $P<B_{DR3}$ to calculate mass ratios and periods. We repeat this 100,000 times for each source, and use the median inferred mass ratio and period as our final values with uncertainties given by 68\% confidence intervals.

The inferred mass ratios and periods using Equations \ref{qfullsolution} and \ref{period} versus the true values are shown in Figures \ref{fig:ruwes_simulation_inferred_q} and \ref{fig:simulation_inferred_q}. We see that the inferred values are well-correlated with true values when the binary contribution to \textit{RUWE} is significant ($\textit{RUWE}_{spec}>2,RUWE_{ast}>1.25$). At short periods (when $\textit{RUWE}_{ast}$ is insignificant), mass ratios are overestimated while at long periods (when $\textit{RUWE}_{spec}$ is small) mass ratios are underestimated, confirming the predictions based on Figure \ref{simulated_beta_zeta_period}. 

We note that because our aim is to identify systems with high mass ratios, we do not attempt to remove sources with underestimated $q$. However, Figure $\ref{simulated_beta_zeta_period}$ shows that a cut on inferred period can be used to remove sources with low values of $\zeta$ and $\beta$ (and therefore sources with underestimated mass ratios) if desired.

Finally, we also note that when the true mass ratio $q$ is very small ($<10^{-2}$), Equation \ref{eqn:sigma_rv} predicts that the radial velocity error contribution from the binary will tend towards 0, in which case the spectroscopic noise should dominate the measurements. As Figure \ref{fig:simulation_inferred_q} shows, this limits our ability to probe binaries with very small mass ratios (e.g. brown dwarfs and exoplanets), since the inferred mass ratios will be overestimated as a result of the noise floor. 

\subsection{Ellipsoidal variation and photometric \textit{UWE}}

Because luminous stars with compact object companion can be tidally distorted such that the light curve shows ellipsoidal modulation, we find it useful to describe the photometric variability via
\begin{equation}
    \textit{UWE}_{phot}=\sqrt{\frac{N_{F}}{N_{F}-1}}\frac{\sigma_{F}}{\sigma_{phot}},
\end{equation}
where $N_{F}$ is the number of observations contributing to the measured flux, $\sigma_{phot}$ is photometric noise, and $\sigma_{F}$ is the standard deviation of the measured flux. Using Equation 78 of \cite{Penoyre_2019}, we calculate the flux variability caused by tides as
\begin{equation}
    \sigma_{F,tides}^2=F^2 \cdot\rm{Var}\Bigg[-\tau\frac{m_2}{m_1}\bigg(\frac{R}{a}\bigg)^3\frac{3\sin^2\theta_v\cos^2\phi_v-1}{(1-e\cos\eta)^3}\Bigg]
\end{equation}
where $R$ is the radius of the primary, $F$ is the measured flux of the primary, $\tau$ is a pre-factor of order unity, and $\rm{Var}[\cdot ]$ denotes the variance.  

To measure photometric \textit{UWE} with our simulated systems, we use a constant photometric error of $\sigma_{phot}=10^{4}$ es$^{-1}$ based on the typical values seen in \textit{Gaia} data (this is discussed in more detail in Appendix \ref{ap:ruwephot}). Figure \ref{fig:RUWE_phot} shows the distribution of $\textit{UWE}_{ast}$ versus $\textit{UWE}_{phot}$ for our simulated systems, and we see virtually no overlap between systems with high astrometric \textit{UWE} and photometric \textit{UWE}, since the orbital separations at which significant ellipsoidal variations occur are too small to produce detectable $\textit{UWE}_{ast}$ signals (recall $\sigma_{b,\theta} \propto a$ from Equation \ref{eqn:sigma_theta}). Because we do not expect candidates selected with our method to be at orbital separations small enough to exhibit ellipsoidal variation, we can instead apply a $\textit{UWE}_{phot}$ cut to \emph{remove} variable sources which we expect to be potential contaminants. 

\section{Agreement with known systems}
\label{sec:comparison}
To test our method we compare two samples of known binary system: the catalogue of non-single star fits to \textit{Gaia} data provided in DR3 \citep{Arenou22}, and an independent sample of spectroscopic binaries from the APOGEE survey \citep{Price-Whelan20}.

\begin{figure*}
\centering
\includegraphics[width=0.98\textwidth]{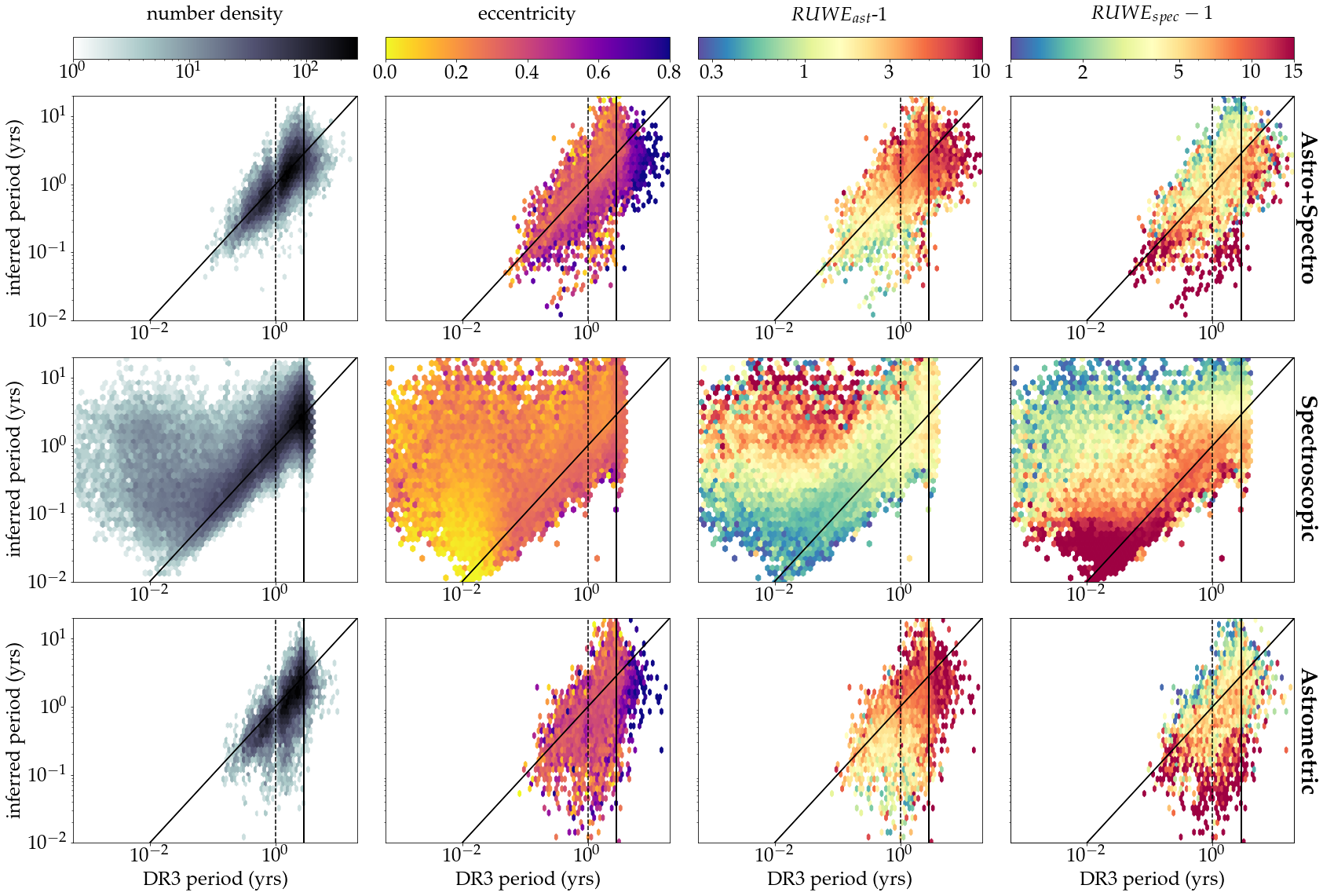}
\caption{True periods for non-single sources presented in the DR3 \code{nss\_two\_body\_orbit} table (with $\textit{RUWE}_{ast}>1.25$ and $\textit{RUWE}_{spec}>2)$) versus those inferred by our method. Vertical lines are drawn at 1 year (dashed) and 34 months (solid). Rows are sorted by  \code{nss\_solution\_type}, with \code{AstroSpectroSB1} (binaries identified with astrometric and spectroscopic data), \code{SB1} (identified with spectroscopic data only), and \code{Orbital} (identified with astrometric data only) solutions for the first (20,831 sources), second (56,527 sources), and third (10,697 sources) rows respectively.}
\label{fig:DR3_periods}
\end{figure*}

\begin{figure}
\centering
\includegraphics[width=0.4\textwidth]{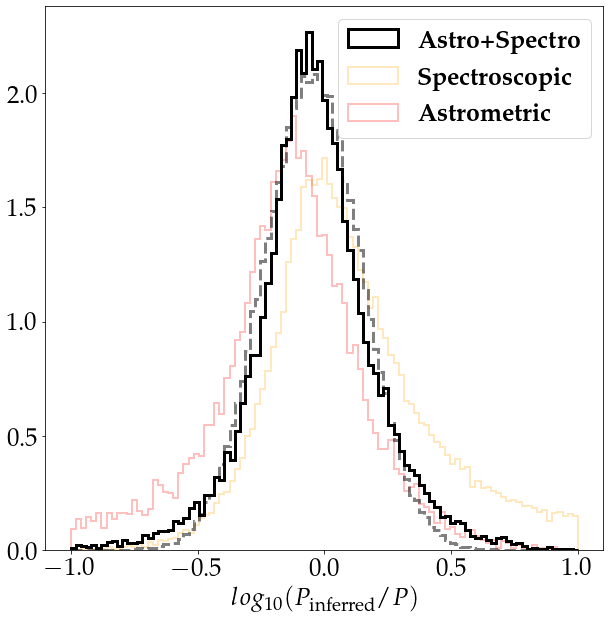}
\caption{Distribution of our inferred periods over measured periods for binaries with combined astrometric and spectroscopic solutions (black), spectroscopic solutions only (yellow), and astrometric solutions only (red) presented in the DR3 \code{nss\_two\_body\_orbit} table. A normal distribution $\pazocal{N}(-0.05,0.2)$ fitted to the distribution for binaries with combined astrometric and spectroscopic solutions is shown with the grey dashed curve.}
\label{fig:periods_scatter}
\end{figure}

\subsection{\textit{Gaia}'s non-single star catalogue}
To begin we process and clean the \textit{Gaia} DR3 Radial Velocity Spectrometer (RVS) sample through a procedure detailed in Appendix \ref{ap:gaia_error}. Our resulting sample contains \rvmethodused sources out of the \fullrvssample sources in DR3 with radial velocity measurements. To calculate the excess radial velocity scatter, $\sigma_{b,vr}$, potentially caused by a binary, we have to find and remove the inherent measurement error $\sigma_{spec}$, which we have calculated from the data as a function of colour and magnitude (see Appendix \ref{ap:ruwespec}). In order to obtain $\sigma_{b,\theta}$, we invert Equation \ref{eqn:RUWE}, substituting  $\sigma_{ast}$ with the robust estimate of the standard deviations of the post-astrometric-fit residuals for along-scan angle measurements as a function of $G$ magnitude, which is provided in \citet{Lindegren21} (solid blue curve of their Figure A.1).
In reality, $\sigma_{ast}$ may be slightly different for individual sources
and as a function of time.

We cross-match our sample to the DR3 \code{nss\_two\_body\_orbit} table which contains orbital parameters for 443,205  binaries including astrometric binaries (identified via their poor fit to the single star astrometric model), spectroscopic binaries (identified based on the spectra collected by \textit{Gaia}), and binaries with combined solutions from astrometric and spectroscopic data \citep{Halbwachs_2022,Holl_2022,Arenou22,Gaia_DR3_Contents}. Since our method only works for sources with significant spectroscopic and astrometric errors, we apply additional cuts of $\textit{RUWE}_{spec}>2$ and $\textit{RUWE}_{ast}>1.25$, leaving 88,606 sources in common. 


We compare our inferred periods in our RVS sample to those measured by \textit{Gaia} in Figure \ref{fig:DR3_periods}. For binaries with combined astrometric and spectroscopic orbital solutions, we see good agreement between inferred periods and true periods. Figure \ref{fig:periods_scatter} shows the log of our inferred periods over the true periods for these binaries. We fit a normal distribution $\pazocal{N}(\mu=-0.05,\sigma=0.2)$, indicating that we typically expect our inferred periods to be off by up to a factor of 1.5.

For systems with only astrometrically or spectroscopically measured orbits we see good agreement for the majority of sources but also significant populations of those with periods which do not agree. In the case of spectroscopic binaries we see an additional population of short period  sources with much higher inferred periods. These sources also have significant $\textit{RUWE}_{ast}$, which we would expect to be negligible for sources with orbital periods on the order of days. As predicted in Section \ref{sec:triples}, this potentially indicates the presence of an additional outer binary with a larger orbital period contributing to the astrometric error and resulting in a larger inferred period than the period measured by \textit{Gaia} (which may be the period of an inner orbit). We see the inverse effect with astrometric binaries, in which the inferred period is lower than the period measured by \textit{Gaia} for longer period ($\sim $ 1 year) sources with very high spectroscopic ruwes. This suggests that, if this particular population of sources are mostly triples, \textit{Gaia} could be measuring the outer orbital periods, while our inferred period is larger a result of the high spectroscopic error due to the inner binary.

\subsection{APOGEE}
We also test our method by cross-matching our $RVS$ sub-sample to the gold binary catalogue presented in \cite{Price-Whelan20}, which provides measured periods and mass ratios using spectroscopic data from APOGEE Data Release 16. We find that 319 of the 1,032 sources in their gold binary catalogue exist in our $RVS$ sample.

\begin{figure*}
\centering
\includegraphics[width=0.98\textwidth]{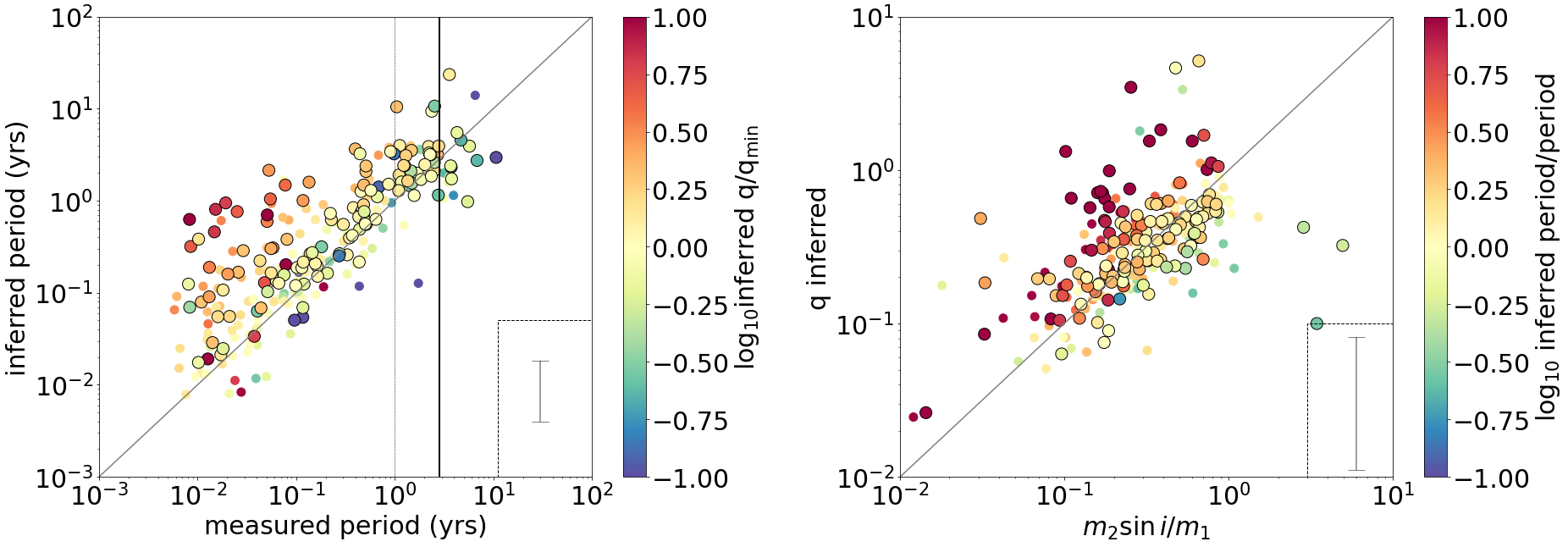}
\caption{Left: true periods for the gold binary catalogue presented in \citet{Price-Whelan20} versus those inferred by our method. Vertical lines are drawn at 1 year and 34 months. Points are colour-coded by inferred versus measured minimum mass ratios. Sources which pass  $\textit{RUWE}_{ast}>1.25$ and $\textit{RUWE}_{spec}>2$ cuts are outlined. Right: measured minimum mass ratios versus inferred mass ratios colour-coded by the ratio of inferred period to true period. Characteristic (median) error bars are shown in inset panels of both plots.}
\label{fig:apogee}
\end{figure*}

Figure \ref{fig:apogee} shows the inferred periods and mass ratios for binaries in the APOGEE gold binary catalogue versus their measured values. We compare our inferred $q$ to the 50$^{th}$ percentile value of $m_{2}\sin i$ (minimum secondary mass) divided by the primary mass given in the catalogue. 

We see good agreement between inferred periods and true periods, though the scatter is large (about an order of magnitude). Sources that have overestimated mass ratios also tend to have overestimated periods. The periods are on average higher than the measured values, which is consistent with what we would expect for most binaries with $l\neq 0$ as discussed in Section \ref{sec:Parameters_from_measured_errors} and Figure \ref{fig:l_v_q_massratio}. 

There is also reasonable agreement between the APOGEE mass ratios and the values we infer. Since the mass ratios from the APOGEE catalogue are calculated from the minimum secondary mass $m_2 \sin i$, our mass ratios are on average larger than the mass ratios from the catalogue. 

It is interesting to note that even for systems without significant \textit{RUWE}, the period and mass ratios are reasonably well-estimated. Their (marginal) astrometric and spectroscopic errors are still measures of the binary properties, but without them being externally selected via APOGEE we would not be able to pick them out as significantly different from single stars.

\section{Candidate \textit{Gaia} compact objects companions and close triples}
\label{sec:data}

\begin{figure}
\centering
\includegraphics[width=0.4\textwidth]{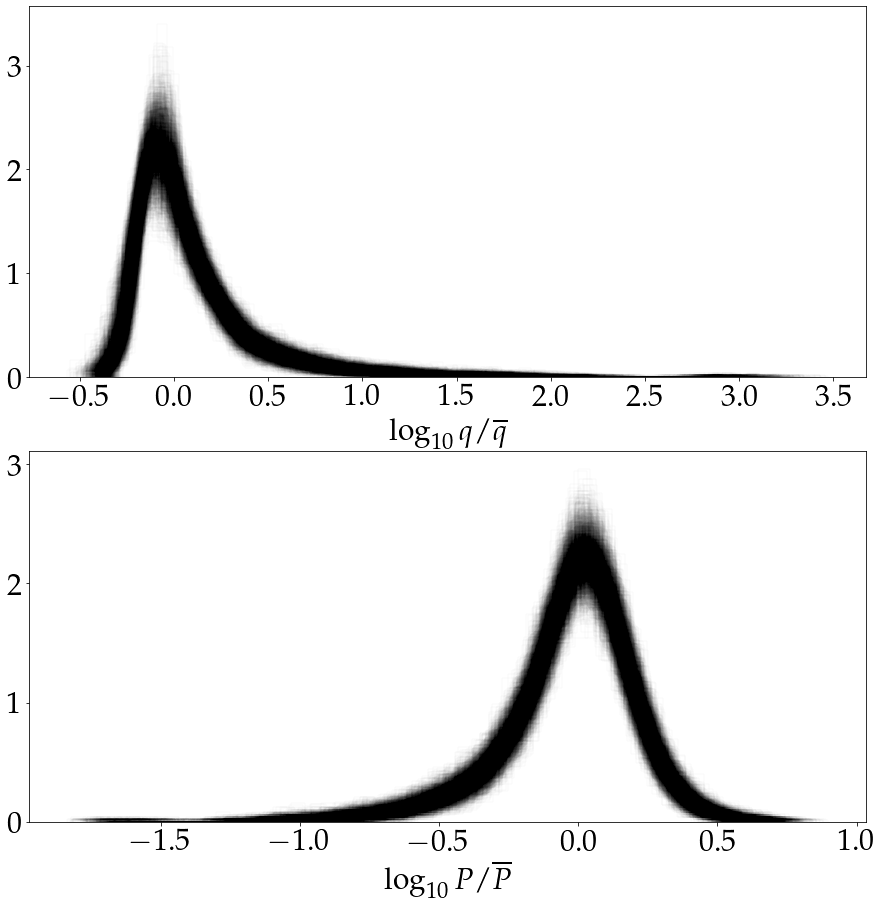}
\caption{Distribution of inferred mass ratios (top) and periods (bottom) normalised by median values for all stars in our bronze sample (\bronzelist sources).}
\label{q_p_err_dist}
\end{figure}

\begin{figure*}
\centering
\includegraphics[width=0.8\textwidth]{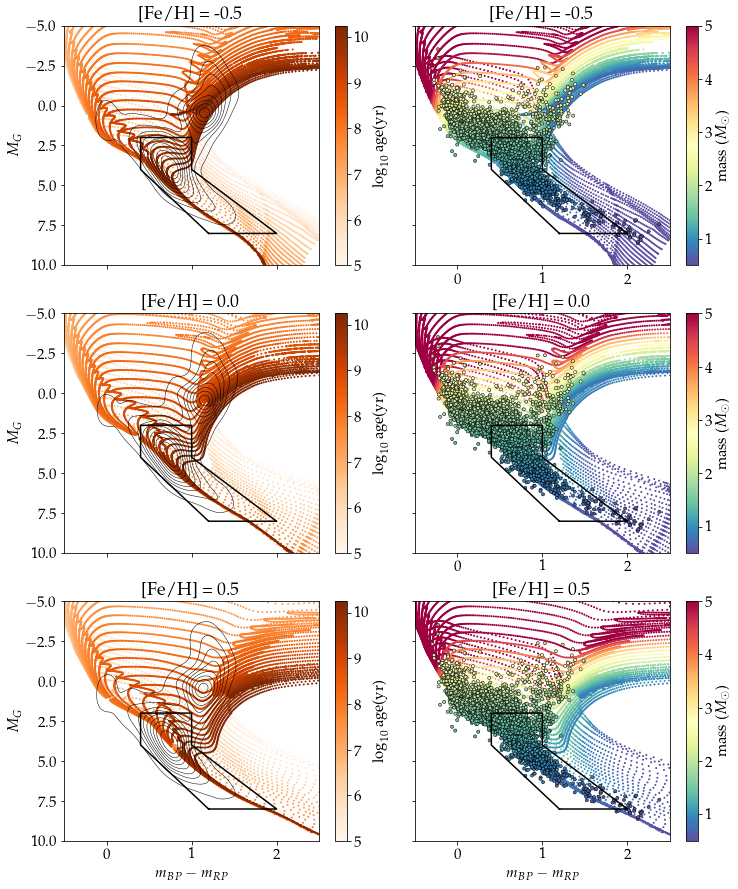}
\caption{Isochrones taken from \citet{Choi_2016} \citet{Dotter_2016} and  \citet{Paxton_2015} plotted on the HRD, assuming sub-solar (top panels) solar (middle) and super-solar (bottom) metallicities. Density contours of our RVS sample on the HRD are plotted in the foreground on the left. In the right column, our bronze sample is over-plotted, colour-coded by inferred primary masses. The main-sequence mask is outlined in black and defined by the points $m_{BP}-m_{RP}=[1.2,2,1,1,0.4,0.4]$ and $M_G= [8,8,4,2,2,4]$.}
\label{fig:HRD}
\end{figure*}

\begin{figure*}
\centering
\includegraphics[width=0.98\textwidth]{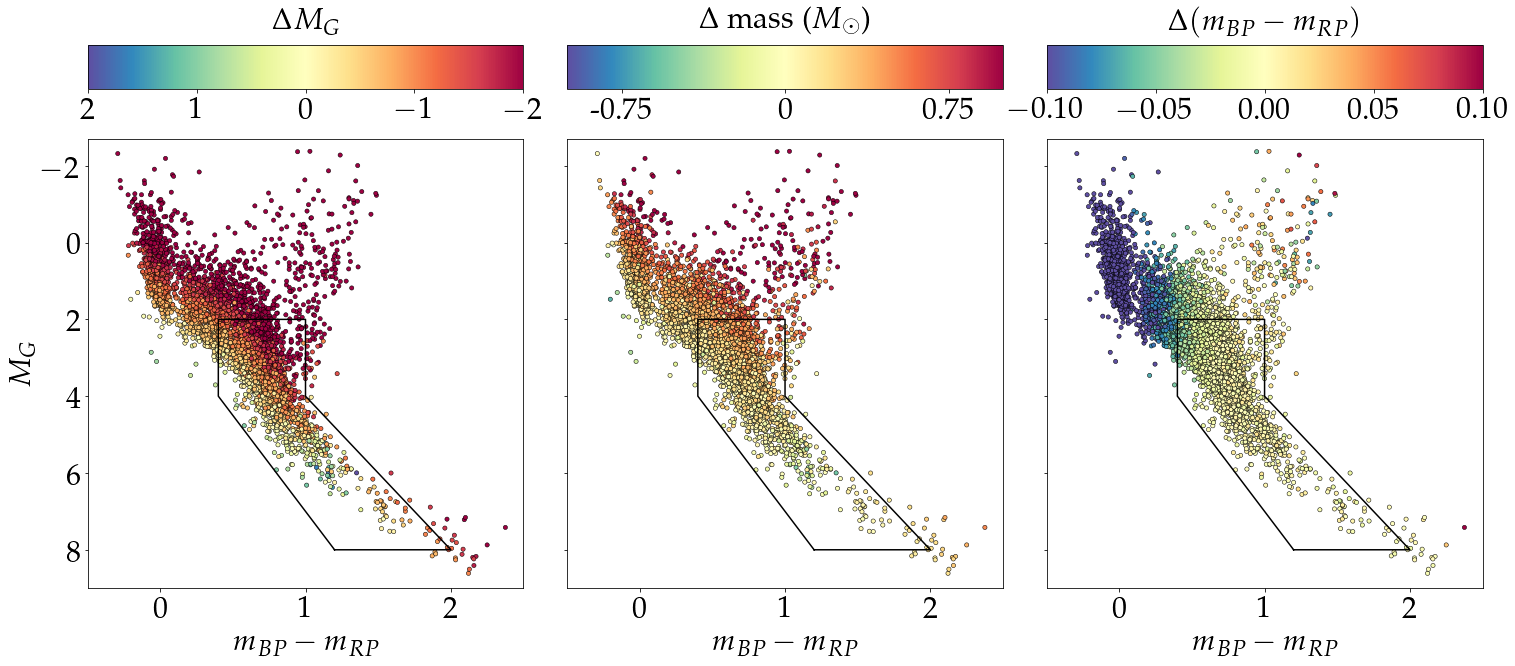}
\caption{\emph{Left}: Distribution of the difference between the absolute $G$ magnitude of our bronze candidates and the mean absolute $G$ magnitude of the main sequence isochrones obtained from  \citet{Choi_2016,Dotter_2016,Paxton_2015}; i.e., the average height each source lies above the main sequence on the HRD. \emph{Middle}: distribution of the difference in the inferred mass of the primary and the average mass of the corresponding source on each isochrone track; sources that are bright and on the RGB tend to have overestimated primary masses when using the main-sequence mass-magnitude relation in \citet{Charalambos_2019}. \emph{Right}: Same as the left and middle plots, except comparing the difference in colour: candidates with $m_{BP}-m_{RP}<0.5$ lie significantly to the left of the corresponding source on each isochrone track. }
\label{fig:delta_mag_candidates_HRD}
\end{figure*}

\begin{figure*}
\centering
\includegraphics[width=0.98\textwidth]{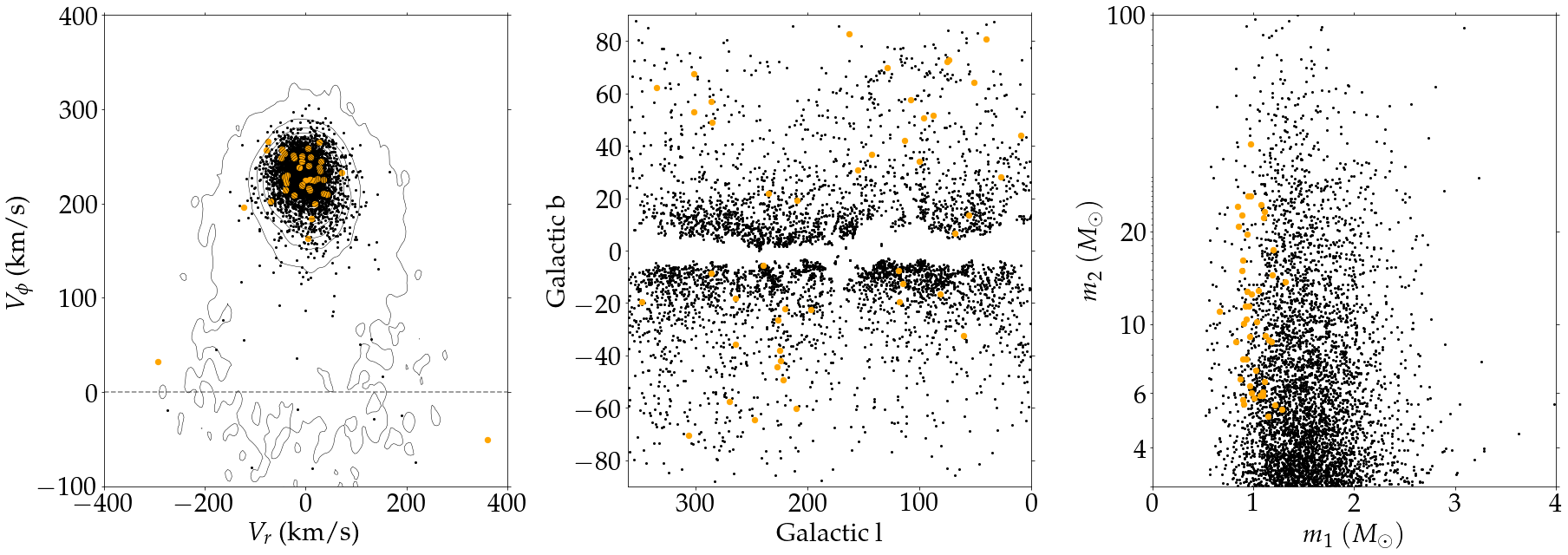}
\caption{Properties of the bronze (gold) candidate sample shown as black (golden) points. {\it Left:} Azimuthal and radial velocity components in galacto-centric spherical polars are shown for systems with heliocentric distances $<1.5$ kpc. Grey-scale density contours correspond to the stars in the {\it Gaia} DR3 RVS sample with heliocentric distances $<1.5$ kpc. CO candidates in our sample sit firmly within the thin disc portion of the distribution. {\it Middle:} Distribution of the candidates in heliocentric Galactic coordinates $l$ and $b$. Regions with high extinction and high source density typically towards the Galactic centre are avoided. {\it Right:} Inferred secondary mass as a function of inferred primary mass. Reassuringly, no obvious correlation is visible.}
\label{fig:bh_candidates_kinematic}
\end{figure*}

\begin{figure*}
\centering
\includegraphics[width=0.98\textwidth]{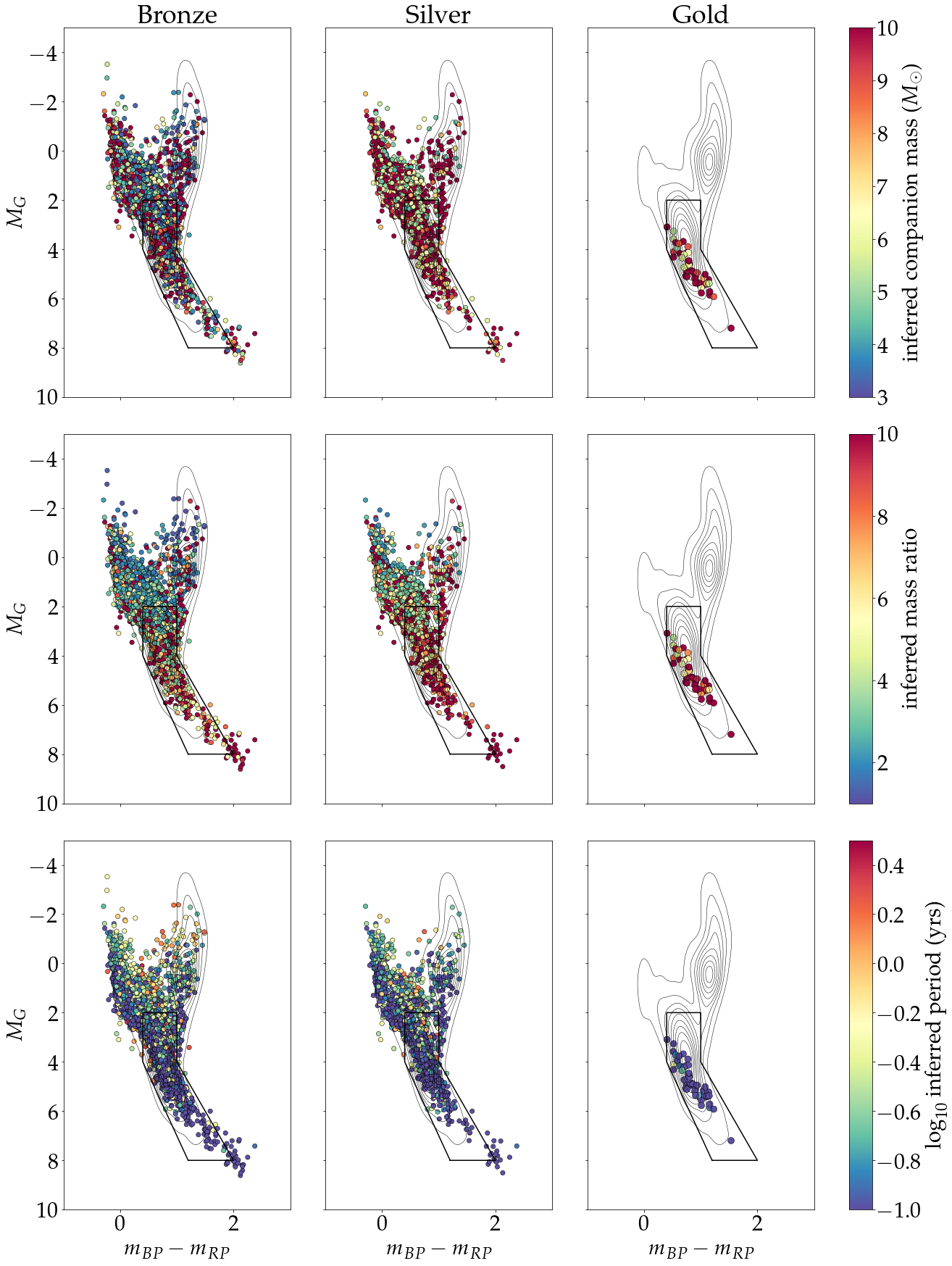}
\caption{Distribution of bronze (left column) silver (middle) and gold (right) plated candidates on the HRD colour-coded by inferred companion mass (top row), mass ratio (middle) and period (bottom). Grey-scale density contours correspond to our full RVS sample.}
\label{fig:HRD_Candidates}
\end{figure*}

\begin{figure*}
\centering
\includegraphics[width=0.98\textwidth]{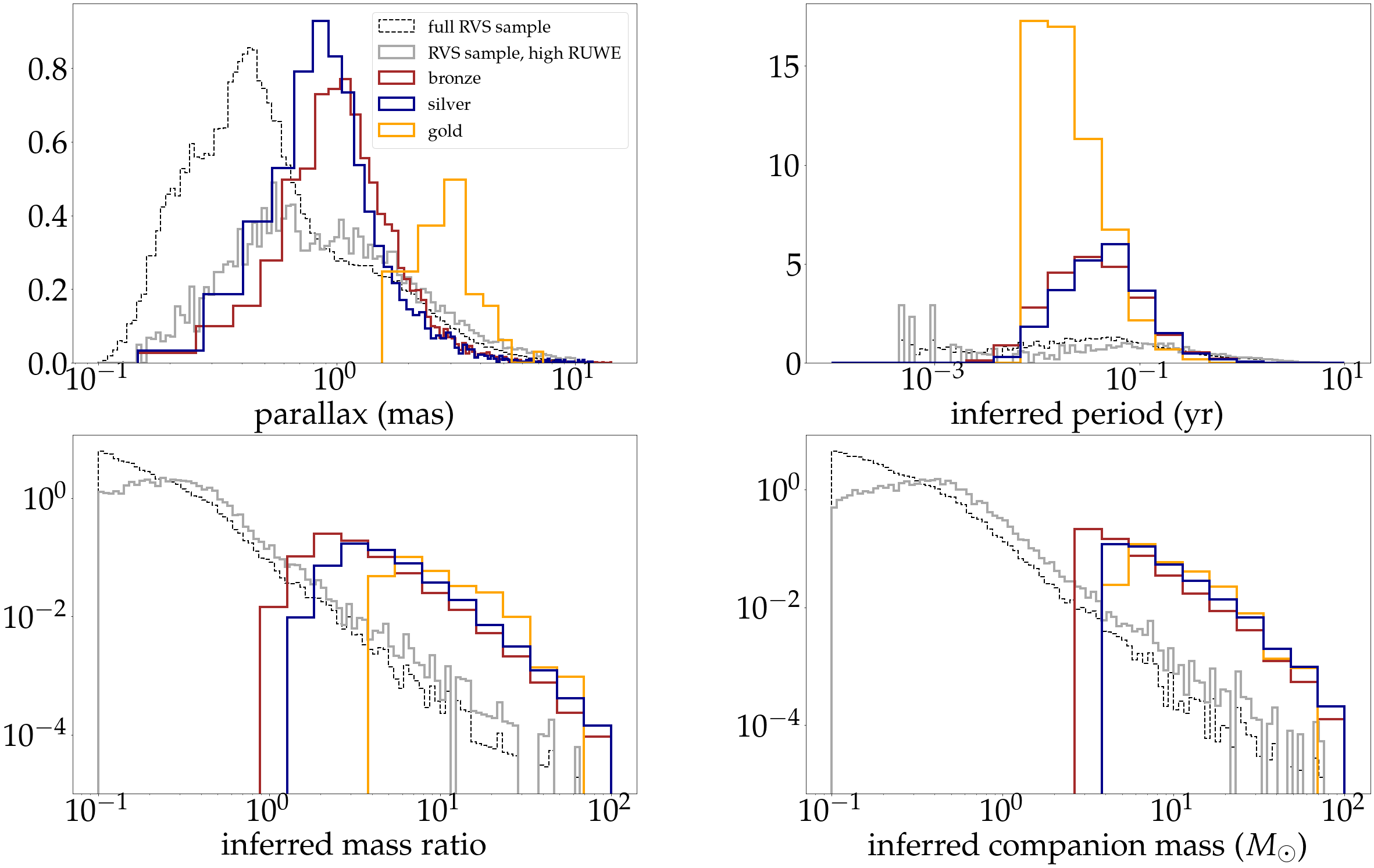}
\caption{Normalised distributions of parallaxes, inferred periods, inferred mass ratios, and inferred companion masses for bronze, silver, and gold-plated samples, as well as our full RVS sample (dashed black line) and sources in our RVS sample with $\textit{RUWE}_{ast}>1.25$, $\textit{RUWE}_{spec}>2$ (solid grey line).}
\label{fig:Hists_Candidates}
\end{figure*}

\subsection{Candidate systems}
Here we describe the selection procedure we have adopted for our dark-remnant binary candidates. In addition to the cuts described in Appendix \ref{RVS_sample}, the following cuts are applied to the \textit{Gaia} DR3 $RVS$ dataset to obtain our initial bronze sample of candidates:
\begin{itemize}
      \item $\code{RUWE}_{spec} > 2 $
      \item $\code{RUWE}_{ast} > 1.25  $
      \item $\code{ipd\_frac\_multi\_peak} = 0$ 
      \item $N_{4\arcsec}=1  $
      \item $q_{\rm{{inferred}}}>1 $ 
     \item  $m_{2,\,\rm{{inferred}}}>3M_\odot  $
\end{itemize}
The cuts on $\textit{RUWE}_{spec}$ and $\textit{RUWE}_{ast}$ are used to separate binaries from single sources and obtain systems with significant spectroscopic and astrometric errors. The cut on $\code{ipd\_frac\_multi\_peak}$ is meant to remove binaries with two partially resolved luminous components. 
Additionally, we require no neighbouring stars within 4\arcsec ($N_{4\arcsec}=1$) to avoid contamination due to blending. This leaves 147,475 sources.

To infer the mass of the primary based on its absolute \textit{Gaia} $G$ magnitude, we use the main-sequence mass-magnitude relation presented in \cite{Charalambos_2019}. We subsequently place a cut of $q_{\rm{{inferred}}}>1$ and $m_{2,\rm{inferred}}>3M_\odot$, leaving \bronzelist sources which constitute our bronze sample. 

As mentioned in Section \ref{sec:simulated_systems}, because $\beta$ and $\zeta$ are unknown values, we use the Monte Carlo method of error propagation to determine uncertainties on inferred mass ratios and periods by drawing from the distributions of $\zeta$ and $\beta$ in Figure \ref{fig:beta_zeta_hist} 100,000 times. We note that for all systems, we assume the same underlying distribution of $\zeta$ and $\beta$. Figure \ref{q_p_err_dist} shows the distribution of mass ratios and periods (100,000 for each source) normalised by median values for all sources in the bronze catalogue.

For our gold and silver samples, we also require that the 68\% confidence intervals of $q_{\rm{{inferred}}}$ and $m_{2,\rm{inferred}}$ lie above 1 and $3M_\odot$:
\begin{itemize}
  \item $q_{\rm{{inferred}}}-\sigma_{q,\rm{{inferred}}}>1$ 
   \item $m_{2,\rm{inferred}}-\sigma_{m2,\rm{{inferred}}}>3M_\odot $ 
\end{itemize}
which leaves \silverlist sources for our silver sample. For our gold sample, we apply the following additional cuts:
\begin{itemize}
   \item $N_{8\arcsec}=1 $
    \item $\textit{RUWE}_{phot}<2 $
   \item $\code{MSmask==True}$
   \item $ \Delta M_G \leq 0 $
\end{itemize}
We apply a stricter cut on crowding by removing sources with neighbours within 8". Additionally, because we do not expect our candidates to be at orbital separations small enough to exhibit ellipsoidal variation, we remove variable sources with $\textit{RUWE}_{phot}>2$ (see Appendix \ref{ap:ruwephot}). This cut is important as unresolved triples may be a significant contamination of our sample, and our analysis (which assumes astrometric and photometric noise come from the same orbit) will break down if the inner pair causes significant radial velocities whilst the outer dominates the astrometric motion. As shown in Figure \ref{fig:RUWE_phot} there is almost no overlap between binary systems with significant $\textit{RUWE}_{phot}$ and those with significant $\textit{RUWE}_{ast}$. This cut may also remove some Variability Induced Movers \citep{Wielen96} where the astrometric motion comes from a wide binary in which one (or both) of the components is varying in brightness, causing the photocentre to move over time. 

We also require all sources to lie within the black main sequence mask shown in Figure \ref{fig:HRD} ($\code{MSmask==True}$), which compares the distribution of remaining candidates on the HR diagram to MIST isochrones\footnote{We use the synthetic photometry for \textit{Gaia} colour-bands based on \cite{Riello21} provided by MIST isochrones in our HRD plots.} \citep{Paxton_2015, Dotter_2016, Choi_2016} with sub-solar, solar, and super-solar metallicities. Because the discrepancy between theoretical masses calculated by MIST and the primary masses we calculated using the mass-magnitude relation given by \cite{Charalambos_2019} is large for sources that lie significantly outside the main sequence, we remove all sources that do not lie within the black mask shown in Figure \ref{fig:HRD}. 

Finally, our fourth cut requires that candidates lie on or below the main sequence of theoretical isochrones ($\Delta M_G \leq 0$), since a position significantly above the main-sequence indicate the presence of a luminous companion. To do this, we first compute the metallicities of our candidates with an algorithm trained on the BP/RP spectra of APOGEE stars\footnote{\textit{Gaia} also contains metallicities computed from DR3 RVS spectra (\code{fem\_gspspec} and \code{mh\_gspspec}) in the \code{astrophysical\_parameters} table. However, we have computed our own calibration of [Fe/H] since \code{fem\_gspspec} and \code{mh\_gspspec} are provided for only a small ($<10\%$) subset of our candidate list.}. We selected $\sim$ 640,000 APOGEE DR17 stars with low extinction E[B-V]<0.1,  measured by APOGEE [Fe/H] abundances and BP/RP spectra. We took an array of BP/RP coefficients for the stars (two 55 element vectors for each star) and divided them by the G magnitude fluxes.
We then fit a random forest regressor (from the \code{sklearn} package) between the 110 element long vectors of normalised BP/RP coefficients and [Fe/H]
to predict metallicity for any star with a BP/RP spectrum. The typical accuracy (based on 16\%/84\% percentiles of residuals)
achieved on a hold-out data-set was 0.08 dex.

For each source, we compare the candidates' HRD positions with the isochrones of the corresponding metallicity. We do this probabilistically by assuming a log-uniform distribution of ages
\begin{equation}
    \rm{age}\sim 10^{\pazocal{U}(7,10)} (yrs)
\end{equation}
and computing average difference in absolute magnitude $\Delta M_G$ from the star’s position in HRD to the corresponding isochrone over 10,000 iterations. Figure \ref{fig:delta_mag_candidates_HRD} shows the distribution $\Delta M_G$ for our sources on the HRD. We remove sources which on average have excess flux compared to the corresponding isochrone of the same metallicity ($\Delta M_G<0$), leaving \goldlist sources for our gold sample.

As shown in Figure \ref{fig:bh_candidates_kinematic} our sources lie mostly within the (thin) disk of the Milky Way (spatially and kinematically) and thus likely contain a range of metallicities. However, we can see in Figure \ref{fig:HRD} that the broad distribution of our sources is consistent with [Fe/H] somewhere between 0.0 and +0.5, suggesting a young stellar population. Our candidates are also more densely distributed at low galactic latitudes, which is unsurprising given that this region of the sky contains more stars.

Finally, in Figures \ref{fig:HRD_Candidates} and \ref{fig:Hists_Candidates} we show some simple properties of our candidates. The former shows the companion mass, mass ratio and periods as a function of position on the HR diagram. Only the mass ratio shows a notable behaviour, with smaller mass ratios for brighter (more massive) systems. This is mostly due to our cut on companion mass being, in general, more stringent than our cut on mass ratio. Lower mass ratio systems are likely more common, but given the above cuts are only included in our candidate list for massive primaries.

Whilst there are a few Sun-like systems below the Red Giant turnoff (at $M_G \approx 4$), the majority are brighter stars--particularly on the Young Main Sequence (especially after we remove Giants as we believe our estimates for their masses are unreliable). This makes sense from an observational perspective--we expect MS+CO systems to be relatively rare and thus we need to look to large distances before we see a significant number, hence preferring brighter systems for which accurate astrometry and spectroscopy can still be performed. 

There is similarly a plausible stellar evolution argument--we expect both components of a binary to have similar masses \citep{Moe17}. Any compact object in such a binary was likely once the brighter and more massive primary, such that it exhausted its fusible material first and collapsed. In order to form a CO, its mass must have been high, likely more than 10 $M_\odot$ \citep{Heger03} and its lifetime relatively short. Thus, we would expect the original secondary (now the visible luminous component) to be similarly young and bright.

Turning to Figure \ref{fig:Hists_Candidates} we show the distribution of the inferred parameters of each catalogue of our candidates and their distances--as well as the properties of the entire RVS sample and the subset with significant \textit{RUWE}. The candidate systems tend to reside at distances of around 1 kpc, with the gold sample in particularly tending to be slightly closer. More distant systems are dimmer and hence the astrometric and spectroscopic error likely becomes too high for significant \textit{RUWE}, whilst the expected rarity of MS+CO systems likely accounts for the dearth of high parallax candidates.

Our candidates have lower inferred periods, matching our assertion that periods can only be reliably inferred below the baseline of the survey. The mass ratios and companions masses are, by design, high and are larger for each more selective subsample. Some systems have extremely high mass ratios and companion masses ($q>10$ and $m_2>10 M_\odot$) though it is hard to gauge if these systems are really this extreme or are just the high-tail end of the distribution of possible $\beta$, $\zeta$ and other (unmodelled) random noise.

We also find that 1129 of our bronze candidates have been flagged as variable sources (based on the \code{phot\_variable\_flag} column of the \code{gaia\_source} table for DR3), of which 670 are classified as eclipsing binaries, 269 are classified as $\delta$ Scuti/$\gamma$ Doradus/Phoenicis stars (\code{DSCT|GDOR|SXPHE}), 139 as solar-like stars (\code{SOLAR\_LIKE}), 12 as RR Lyrae, 8 as young stellar objects, 7 as long period variables, 4 as $\alpha$ Canum Venaticorum/(Magnetic) Chemical Peculiar/Rapidly Oscillating Am-type/SX Arietis stars (\code{ACV|CP|MCP|ROAM|ROAP}), 3 as Cepheids, 1 as a $\beta$  Cephei variable, and 1 as a Slowly Pulsating B-star variable in the \textit{Gaia} DR3 \code{vari\_classifier\_result} variability table \citep{Eyer_2022}. The high fraction of eclipsing binaries in our candidate list ($\sim 60\%$) compared to the fraction in the entire \code{vari\_classifier\_result} table (22\% as shown in Figure 4 of \cite{Eyer_2022}) is to be expected, especially given our prediction that the main source of contaminants for compact objects are triples and higher order systems with tight inner binaries (this is discussed further in Section \ref{sec:triples_candidates}). Only 8 of our gold sources are flagged as variables based on  \code{phot\_variable\_flag} (all of which are classified as solar-like stars) which is likely due to our photometric ruwe cut removing variables. 

Additionally, we find that 1031 of our bronze candidates have been flagged in \code{gaia\_source} as non-single-sources (based on \code{non\_single\_star}), of which 690 are classified as astrometric binaries, 190 as spectroscopic, 46 as eclipsing, and the remaining 105 as combinations of the three.

\begin{figure*}
\centering
\includegraphics[width=0.98\textwidth]{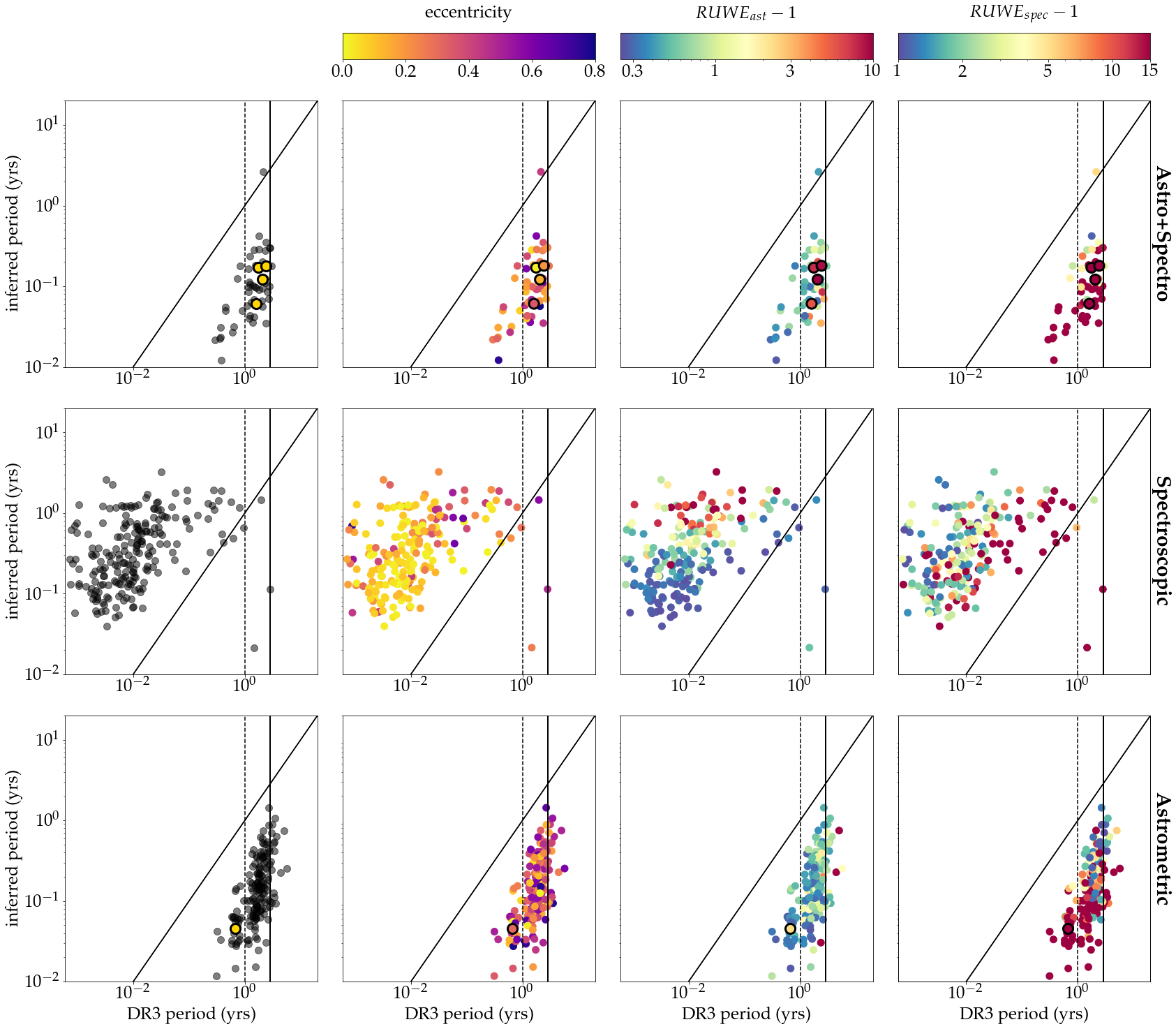}
\caption{Same as Figure \ref{fig:DR3_periods}, except with our bronze candidate sample. Sources in our original gold sample (5) are colour-coded gold in the leftmost column and are given an outline in all other plots.}
\label{fig:DR3_candidates_periods}
\end{figure*}

\subsection{Triples and higher-order multiples in candidate list} 
\label{sec:triples_candidates}
Figure \ref{fig:DR3_candidates_periods} shows our inferred period versus the period measured by \textit{Gaia} for sources in our bronze candidate list that also exist in the \code{nss\_two\_body\_orbit} table (499 out of \bronzelist sources in our bronze candidate list). Unlike Figure \ref{fig:DR3_periods}, we see that the majority of our candidates have periods that are overestimated (if measured as spectroscopic binaries by \textit{Gaia}) or underestimated (if measured as astrometric binaries by \textit{Gaia}), consistent with a potential population of  triples. As discussed in Section \ref{sec:triples}, if  spectroscopic and astrometric errors for higher-order multiples are overestimated as a result of being induced by different orbital periods, the inferred mass ratio will also be large. Since our candidate list of MS-CO objects are selected primarily on the criteria of large mass ratios/secondary masses, it is likely that we are also selecting for triples with overestimated astrometric and spectroscopic errors. 

In our gold sample, we attempt to guard against triples with our cut on photometric variability ($\textit{RUWE}_{phot}<2$). However, Figure \ref{fig:DR3_candidates_periods} also shows that 5 of our initial list of \goldlistinitial gold-plated candidates exist in the \code{nss\_two\_body\_orbit} table; each has a discrepancy between our inferred period and the period measured by \textit{Gaia} that is consistent with triples. While we remove these sources from our final gold candidate list (\goldlist sources), it's likely that triples and higher-order multiples are a major contaminant in even our gold candidate list. 

To estimate the fraction of triples contained in our sample, we cross-match our bronze candidate list with unresolved spectroscopic triples from APOGEE DR17 presented in \cite{Kounkel_2021}. We find that 138 sources in our bronze candidate list are in APOGEE DR17, and 49 contain one of the following flags indicating problematic measurements: \code{STAR\_BAD, TEFF\_BAD, LOGG\_BAD, VERY\_BRIGHT\_NEIGHBOR, LOW\_SNR, PERSIST\_HIGH, PERSIST\_JUMP\_POS, PERSIST\_JUMP\_NEG, SUSPECT\_RV\_COMBINATION}. Of the remaining 89, 8 (9\%) have been classified as spectroscopic triples by \cite{Kounkel_2021}, which we consider to be the minimum triple contamination threshold in our candidate list.

\section{Conclusions}

In this paper, we explore how spectroscopic and astrometric errors can be used to infer the mass ratios and periods of unresolved binary systems. Specifically, we have developed a method for inferring the mass ratios and periods for binaries with very small light ratios ($l \ll 1$), and have also identified the light-ratio regimes in which our method results in overestimated and underestimated values. 

We also show that the projection terms $\zeta$ and $\beta$--which parameterise the dependence of spectroscopic and astrometric errors on viewing angle, period, and eccentricity--are period-independent for binaries with $P<B$. Even for sources in which the viewing angle, eccentricity, and period are unknown, $\zeta$ and $\beta$--which are needed to calculate the period and mass ratio--can still be approximated by assuming $P<B$ and drawing repeatedly from simulated distributions of $\zeta_0$ and $\beta_0$. We show approximately how $\zeta$ and $\beta$ decay with period for $P>B$, and consequently, that binaries with periods larger than the observational baseline have underestimated mass ratios. However, we have also shown that these sources can mostly be removed with \textit{RUWE} cuts.

We construct an analogous statistic to the astrometric renormalised unit weight error ($\textit{RUWE}_{ast}$) for spectroscopic and photometric errors ($\textit{RUWE}_{spec}$ and $\textit{RUWE}_{phot}$, respectively). We find, after quality cuts on $\textit{RUWE}_{spec}>2$ and $\textit{RUWE}_{ast}>1.25$, there is good agreement between our predicted periods and mass ratios to true values with both simulated systems and known binaries in APOGEE. 

We apply our method to \textit{Gaia} DR3 astrometric and radial velocity data in order to obtain a candidate list of luminous stars with massive dark companions. We find \bronzelist sources with inferred mass ratios larger than unity and inferred companion masses larger than $3M_\odot$. We use the Monte-Carlo method of error-propagation to obtain a 68\% confidence interval on the inferred mass ratios and secondary masses by drawing from simulated distributions of $\beta$ and $\zeta$. After removing sources with 68\% confidence intervals below $q=1$ and $m_2=3M_\odot$ and applying  additional quality cuts, we obtain our initial gold candidate list containing \goldlistinitial sources.  

Some degree of contamination, most notably by unresolved triples (and higher multiples) as well as the occasionally purely spurious \textit{Gaia} measurement, is inevitable. Particularly rife for misinterpretation are multiple star systems for which large spectroscopic and astrometric \textit{RUWE}s stem from different pairs with different periods. Our candidates appear to be almost entirely in the disk, and are likely young and metal rich, which is consistent with a MS+CO interpretation. Further probing the nature of these systems will likely require examining them through other astronomical lenses, such as their light curves, spectra and direct imaging. For instance, identifying triples with an inner binary could be done by examining light curves and extracting sources with periods too short to induce a high astrometric error.

Even in cases where these systems do not host compact objects they are still some of the most extreme and thus potentially engaging within the current \textit{Gaia} data. These candidates form an exciting--but preliminary--sample of possible MS+CO binaries. Future \textit{Gaia} data releases will better constrain the properties of these systems, as well as lowering the noise floor and allowing us to select yet more candidates.

\begin{deluxetable}{c|c}
\tabletypesize{\normalsize}
\tablecolumns{8}
\tablewidth{.5\textwidth}
\tablecaption{ Catalogue Column Descriptions \label{table:catalog_column_description}}
\tablehead{\colhead{Parameter} & \colhead{Description}} 
\startdata
\code{q} & inferred mass ratio \\
\code{q\_lower} & lower confidence level (16\%)  \\
\code{q\_upper} & upper confidence level (84\%) \\
\code{secondary\_mass} & inferred mass of secondary $(M_\odot)$ \\
\code{secondary\_mass\_lower} & lower confidence level (16\%)  \\
\code{secondary\_mass\_upper} & upper confidence level (84\%) \\
\code{P} & inferred period (yrs)\\
\code{P\_lower} & lower confidence level (16\%)  \\
\code{P\_upper} & upper confidence level (84\%) \\
\code{plating} & Bronze, Silver, or Gold\\
\code{ruwe\_spec} & spectroscopic RUWE \\
... & all columns from the \textit{Gaia} DR3 \\ 
& \code{gaia\_source} table
\enddata
\tablecomments{Description of the data table containing all candidate sources.}
\end{deluxetable}

\section*{Acknowledgements}

We would like to thank all members of the Cambridge Streams group, in particular Sergey Koposov, as well as Lennart Lindegren, Emily Sandford, Jacob Boerma, Kareem El-Badry, Hans-Walter Rix and Katelyn Breivik for their input and discussions. This paper made use of the Whole Sky Database (wsdb) created by Sergey Koposov and maintained at the Institute of Astronomy, Cambridge by Sergey Koposov, Vasily Belokurov and Wyn Evans with financial support from the Science \& Technology Facilities Council (STFC) and the European Research Council (ERC). This work has made use of data from the European Space Agency (ESA) mission
{\it Gaia} (\url{https://www.cosmos.esa.int/gaia}), processed by the {\it Gaia}
Data Processing and Analysis Consortium (DPAC,
\url{https://www.cosmos.esa.int/web/gaia/dpac/consortium}). Funding for the DPAC
has been provided by national institutions, in particular the institutions
participating in the {\it Gaia} Multilateral Agreement.

\section*{Data Availability}

The bronze, silver and gold candidate lists, as well as some relevant \textit{Gaia} DR3 fields and parameters inferred by our method is freely available at \url{https://zenodo.org/record/6779374}. The \changetwo{descriptions} of column values are given in Table \ref{table:catalog_column_description}.

\bibliographystyle{mnras}
\bibliography{bib}
\bsp

\appendix

\section{Radial velocity error}
\label{ap:rvscatter}

The standard deviation of the line-of-sight velocity is given by
 \begin{equation}
     \sigma_{vr} = \sqrt{ \langle v_r^2\rangle-\langle{v}_{r}\rangle^2}.
 \end{equation}
 The radial velocity of a binary is the vector sum of the radial velocity of the system's centre of mass (which we will take to be constant and denote $v_s$) and the radial velocity due to the orbit (denoted by $v_o(t)$). Then
 \begin{align}
     \langle{v}_{r}\rangle^2 &= (v_s+\langle{v}_o\rangle)^2=v_s^2+2v_s\langle{v}_o\rangle+\langle{v}_o\rangle^2\\
     \langle{v_r^2}\rangle&= \langle{(v_s+{v}_o)^2\rangle}=v_s^2+2v_s\langle{v}_o\rangle+\langle{v_o^2}\rangle
 \end{align}
 Thus
 \begin{equation}
    \label{eqn:std_radial_velocity}
     \sigma_{vr} =\sqrt{ \langle{v_o^2}\rangle-\langle{v}_o\rangle^2}
 \end{equation}
 
We can define a Cartesian coordinate system in which the origin is defined to be the centre of mass of the binary and the orbit is confined to the X-Y plane. Assuming the binary is subject to no external forces, then the position of the primary $\vec{r}_1$ and secondary $\vec{r}_2$ at the true anomaly $\phi$ is given by
  \begin{align}
    \label{eqn:r1}
    \vec{r}_{1} &= r(t)\frac{m_2}{M_{tot}}\begin{bmatrix}
           \cos\phi(t) \\
           \sin\phi(t) \\
           0
         \end{bmatrix} = r(t)\frac{q}{1+q}\begin{bmatrix}
           \cos\phi(t) \\
           \sin\phi(t) \\
           0
         \end{bmatrix}\\
        \label{eqn:r2}
    \vec{r}_{2} &=- r(t)\frac{m_1}{M_{tot}}\begin{bmatrix}
           \cos\phi(t) \\
           \sin\phi(t) \\
           0
         \end{bmatrix} =- r(t)\frac{1}{1+q}\begin{bmatrix}
           \cos\phi(t) \\
           \sin\phi(t) \\
           0
         \end{bmatrix}
  \end{align}
where $m_1$ is the mass of the primary, $m_2$ is the mass of the secondary, $M_{tot}$ is the total mass of the binary, $q\equiv m_1/m_2$, and $r(t)$ denotes the separation between the two stars as a function of time. It is known that the orbital separation of a binary in a Keplerian orbit with eccentricity $e$ and semi-major axis $a$ evolves with time as:
\begin{align}
\label{eqn:eta1}
r(\phi(t))=\frac{a(1-e^2)}{1+e\cos\phi(t)}. 
\end{align}
Equivalently, the orbital separation can be more succinctly expressed as a function of the eccentric anomaly $\eta$ 
\begin{equation}
    r(\eta)=a(1-e\cos \eta)
\end{equation}
where $\eta$ is related to $\phi$ as
\begin{align}
\label{eqn:eta2}
    \cos \phi &= \frac{\cos \eta - e}{1-e \cos \eta}\\
\label{eqn:eta3}
    \sin \phi &= \frac{\sqrt{1-e^2}\sin \eta}{1-e \cos \eta}\\
\end{align}
and the time evolution of the binary follows
\begin{equation}
    \label{eqn:time_evolution}
    t=\frac{P}{2\pi}(\eta-e\sin\eta).
\end{equation}
We can define two viewing angles $\theta_v$ and $\phi_v$ to describe the line of sight vector $\hat{k}$ such that $\theta_v$ is the angle between $\hat{k}$ and $\hat{Z}$ (i.e. $\theta_v=0$ when viewing the binary face-on), and  $\phi_v$ is the angle between $\hat{X}$ (which intersects periapse) and the component of $\hat{k}$ projected onto the X-Y plane. Then $\hat{k}$ can be expressed in the X-Y-Z coordinate system as

\begin{align}
    \label{eqn:k}
     \hat{k} &= \begin{bmatrix}
           \sin\theta_v \cos\phi_v\\
           \sin\theta_v \sin\phi_v \\
           \cos \theta_v
         \end{bmatrix}.
\end{align}

 The radial velocity due to the orbit can be calculated by taking the time derivative of the primary, $\vec{r}_1$, projected onto the line of sight $\hat{k}$:
\begin{equation}
    \label{eqn:v01}
    v_o=\frac{d}{dt}(\vec{r}_1\cdot \hat{k})
\end{equation}
which provides
\begin{align}
    \label{eqn:v0}
    v_0=\frac{2\pi a}{P} \frac{m_2}{m_1+m_2}\frac{\kappa_{sc}\sin\eta-\sqrt{1-e^2}\kappa_{ss}\cos\eta}{1-e\cos\eta}.
\end{align}
Substituting Equation \ref{eqn:v0} into Equation \ref{eqn:std_radial_velocity}, we can re-express the RVS in the form given by Equation \ref{eqn:sigma_rv}.

\subsection{Analytical radial velocity scatter in the short period limit}

Substituting Equation \ref{eqn:v0} into Equation \ref{eqn:std_radial_velocity}, we can re-express the RVS as
 \begin{align}
     \sigma_{vr} 
    &=\frac{m_2}{m_1+m_2}\frac{2\pi a}{P}\sqrt{ \langle{w_0^2}\rangle-\langle w_0\rangle^2}
 \end{align}
 where $w_0\equiv v_0(m_1+m_2)P/(m_2 2\pi a)$. Assuming that the binary is frequently and isotropically scanned, we can express the time average of $w_0$ as
\begin{align}
    \langle{w_0}\rangle&=\frac{1}{t_2-t_1}\int_{t_1}^{t_2}dt \ w_0(t) \\
    &= \frac{P}{2\pi a(t_2-t_1)}\int_{t_1}^{t_2}dt \frac{d}{dt}\frac{\vec{r}_1\cdot \hat{k} }{v_c m_2}(m_1+m_2)\\
    \label{eqn:w0}
 &=\frac{\vec{r}_1(t_2)\cdot \hat{k}-\vec{r}_1(t_1)\cdot \hat{k}}{(t_2-t_1)v_c m_2}(m_1+m_2)\frac{P}{2\pi a}
\end{align}
 
Moving on to the time average of $w_0^2$, we have
\begin{align}
    \langle{w_0^2}\rangle&=\frac{1}{t_2-t_1}\int_{t_1}^{t_2}dt \ w_0^2(t) 
\end{align}

Using Equation \ref{eqn:time_evolution}, we can change integration variables from $t$ to $\eta$:

\begin{align}
    \label{eqn:integral}
    \langle{w_o^2}\rangle=\frac{ P}{8\pi^3(t_2-t_1)}\int_{\eta_0}^{\eta_f}d\eta \ \frac{\bigg( -\kappa_{sc}\sin\eta+\kappa_{ss}\cos\eta \sqrt{1-e^2} \bigg)^2}{1-e\cos\eta}.
\end{align}
In the short-period limit where multiple orbits occur over the observational baseline, we can approximate the time average as the time average over one complete orbit. Then $\langle{w_o}\rangle=0$ and
\begin{align}
    \langle{w_o^2}\rangle&=\frac{ P}{8\pi^3(t_2-t_1)}\int_{\eta_0}^{\eta_0+2\pi}d\eta \ \frac{\bigg( -\kappa_{sc}\sin\eta+\kappa_{ss}\cos\eta \sqrt{1-e^2} \bigg)^2}{1-e\cos\eta}\\
    &=\frac{\kappa_{ss}^2\epsilon(1-\epsilon)}{e^2}-\kappa_{sc}^2\frac{(e^4+e^2(3\epsilon-5)+4(1-\epsilon))}{e^2 (1-\epsilon)^2\epsilon}.
\end{align}
This provides the expression for $\zeta_0$ given in Equation \ref{eqn:zeta0}.

\section{Estimating observational error and \textit{RUWE} for \textit{Gaia} spectroscopy and photometry}
\label{ap:gaia_error}
The precision of the measurements (e.g. $\sigma_{spec}, \sigma_{ast}$ and $\sigma_{phot}$) needs to be known in order to calculate the significance of any extra noise (such as that caused by an unresolved binary). However this is often an unknown quantity which we have to find pragmatically with reference to our observations.

In \textit{Gaia} the precision of most measurements is a function of (at least) the magnitude and colour of the source. We will approach this data preparation step in a manner purposefully similar to the construction of the astrometric renormalised unit weight error ($\textit{RUWE}_{ast}$) as detailed in \citet{LindegrenRUWE}. We will use the pre-computed values of $\textit{RUWE}_{ast}$\footnote{As \textit{Gaia} do not currently publish the value of the astrometric error assumed in order to calculate \texttt{ASTROMETRIC\_CHI2\_AL} we cannot perform the necessary analysis to infer $\sigma_{ast}$ from the data.} and calculate $\sigma_{spec}$ and $\sigma_{phot}$, and construct an equivalent measure of the significance, $\textit{RUWE}_{spec}$ and $\textit{RUWE}_{phot}$, here.

\subsection{Selecting the data}
\label{RVS_sample}

We will start from the \textit{Gaia} DR3 catalogue and we'll use the shorthand:
\begin{equation}
RVE=\mathtt{RADIAL\_VELOCITY\_ERROR}
\end{equation}
and
\begin{equation}
RVN=\mathtt{RV\_NB\_TRANSITS}.
\end{equation}

We select sources for which
\begin{itemize}
  \item $RVE>0$
  \item $RVN\geq3$
  \item $\textit{RUWE}_{ast}>0$
  \item $\varpi/\sigma_{\varpi}>10$
  \item \code{rv\_method\_used}=1
  \item $N_{2"}$=1
  \item $m_G$ is finite
  \item $m_{BP}-m_{RP}$ is finite
\end{itemize}
where $N_{2"}$ is the number of \textit{Gaia} sources within 2 arcseconds.
The first cut just ensures that the source is in the RVS sample, and the second that a median and variance of spectroscopic measurements can be meaningfully measured. Cutting on $\textit{RUWE}_{ast}$ ensures that the source has a 5-parameter astrometric solution, and correspondingly that the solution is based on a sufficient number of visibility periods. The cut on parallax over error will limit us to closer sources, where the parallax is relatively reliable, and cut out any negative parallax solutions. Two methods are used for extracting radial velocities from \textit{Gaia's} observations, and we select only those with median and standard deviations calculated using their typical statistical definition (whereas the second method, mostly used on dimmer sources, uses the cross-correlation functions). The cut on nearby neighbours reduces the chance of blending of sources which are close on sky but at different distances biasing our measurements. Finally finite colours and apparent magnitudes are necessary for the following analysis.

This leaves 6,023,733 sources out of the full  33,812,183 in the \textit{Gaia} DR3 radial velocity catalogue. The largest cut is that on the method used for calculating radial velocities. For bright sources (\code{rv\_method\_used=1}), individual measurements of the positions of the spectral lines are taken for each observation, from which a median and a variance can then be measured. For dim sources (\code{rv\_method\_used=2}), the spectra from all observations are stacked to increase the signal to noise. In this case, the error is estimated from the cross-correlation function and is related to the width of the stacked spectrum. The extra noise introduced by a binary scales differently in the two cases, and the treatment presented in this paper--which requires calculating the variance to translate errors into physical parameters--is only directly relevant when \code{rv\_method\_used=1}.

\subsection{Spectroscopic error}
\label{ap:ruwespec}

\begin{figure}
\includegraphics[width=0.47\textwidth]{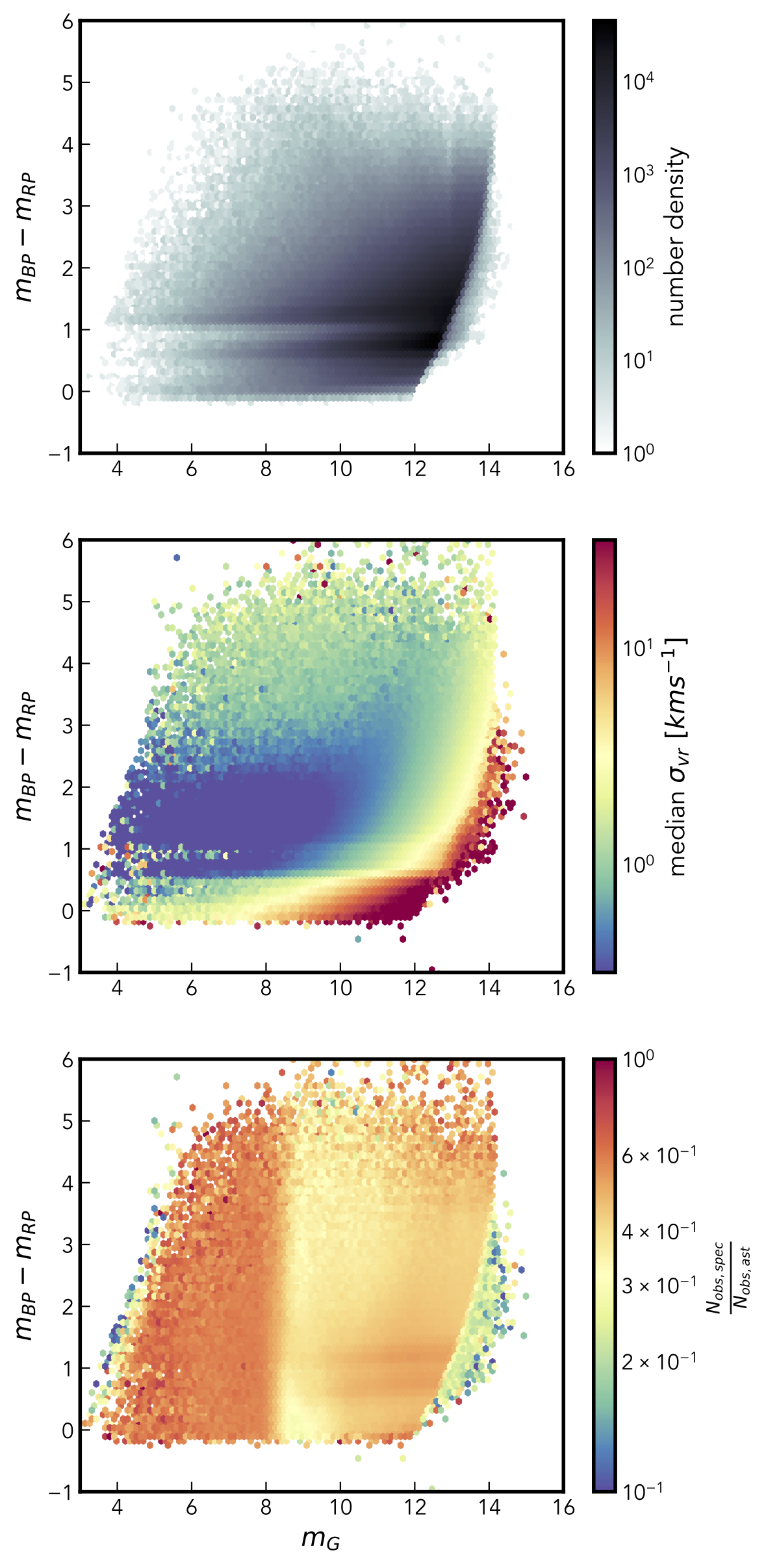}
\caption{Top: Distribution of sources in \textit{Gaia} DR3 RVS sample which pass our selection criteria, as a function of apparent magnitude and colour. Middle: Standard deviation of spectroscopic measurements, $\sigma_{vr}$, from which we can infer a modal $\sigma_{spec}$ which sets the mode of $\textit{RUWE}_{spec}$ to be approximately 1. Bottom: The number of spectroscopic measurements (equal to $RVN$) compared to the number of transits used to compute the astrometric solution ($\code{astrometric\_matched\_transits}$). Overall roughly 1 in 2 observations result in recorded value of radial velocity.}
\label{fig:sigmaspec_bprp_g}
\end{figure}

\begin{figure*}
\includegraphics[width=0.98\textwidth]{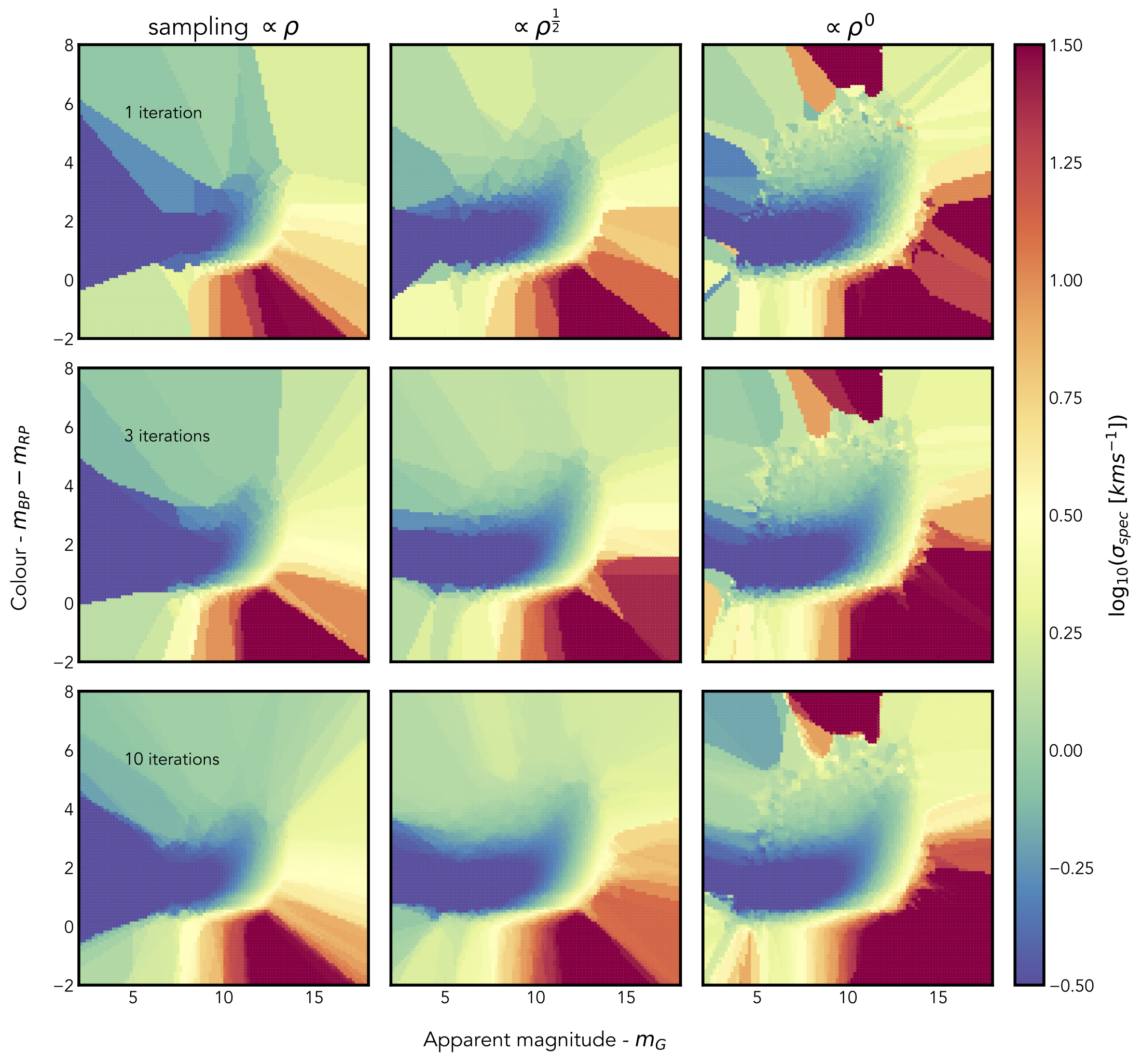}
\caption{Our estimate of the spectroscopic error, as calculated from the value needed to shift the 41$^{st}$ percentile of observed \textit{RUWE}s to 1. Each column shows a different weighting of the selection function the subset of (1000) random points, corresponding to $n=0, -\frac{1}{2}$ and $-1$ respectively (where $n$ is the exponential in Equation \ref{eqn:density_weight}). Each row is a different number of iterations of the method, from which the final value used for each grid point is the median of all iterations. The bottom middle panel corresponds to the inferred errors used through the rest of this work and from which $\textit{RUWE}_{spec}$ is also calculated. As the grid (of 100 by 100) points is relatively coarse some edges, especially for low numbers of iterations, appear jagged.}
\label{fig:renormalised_spec_grid}
\end{figure*}

\begin{figure}
\includegraphics[width=0.49\textwidth]{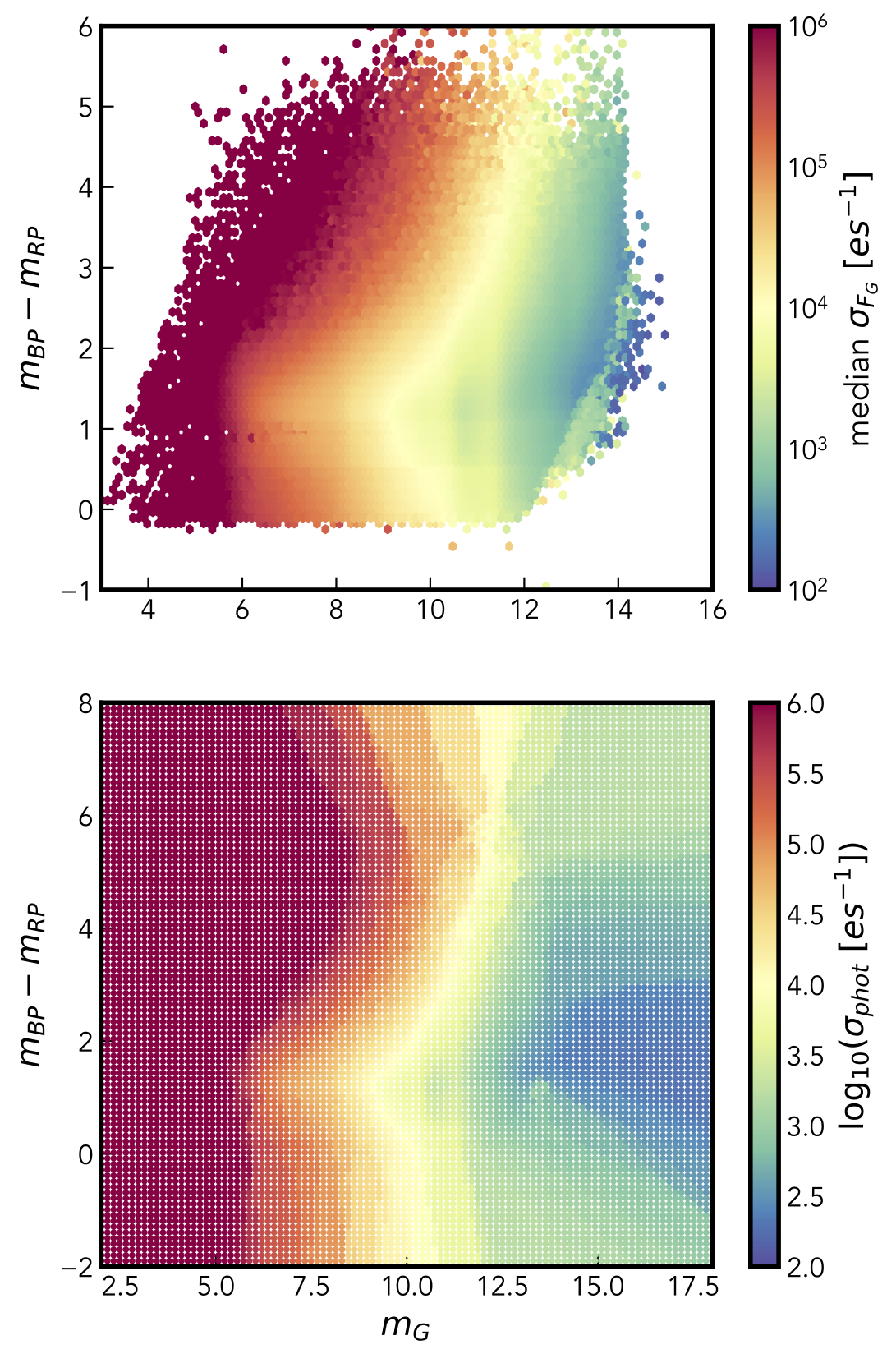}
\caption{Similar to Figure \ref{fig:sigmaspec_bprp_g} but now showing photometric variability (as estimated from \textit{Gaia}'s G band flux). The top panel shows the distribution of the standard deviation of flux, whilst the bottom shows our grid of estimated values.}
\label{fig:sigmaphot_bprp_g}
\end{figure}

There are a few transformations on the quoted \textit{Gaia} data needed before we have the equivalent of a variance of measured velocities. Firstly we must subtract the noise floor of 0.11 km/s added to give the error on the median radial velocity:
\begin{equation}
\sigma_{\overline{v_r}}=\sqrt{RVE^2 - 0.11^2}
\end{equation}
which we can translate back to a standard deviation via
\begin{equation}
\sigma_{vr}=\sqrt{\frac{2 RVN}{\pi}} \sigma_{\overline{v_r}}
\end{equation}
where the factor of $\sqrt{\frac{\pi}{2}}$ comes from the extra variance introduced by using the median rather than the mean of $v_r$.

Figure \ref{fig:sigmaspec_bprp_g} shows the distribution of sources in magnitude and colour, their median standard deviation and the fraction of scans for which  a spectroscopic measurement is made. Understandably the brightest sources can be measured with the lowest errors, though there is a broad range of magnitudes (up to 13 or 14) for which the precision is high, after which the standard deviation increases rapidly due to Poisson noise on a low number of photons. Additionally, we see a strong dependence on temperature, with hotter
stars at a fixed magnitude having larger errors as a result of line-broadening.

We also see strong trends in which sources get more spectroscopic measurements - below $\sim$8$^{th}$ magnitude roughly two thirds of scans result in a recording, likely corresponding to the relative geometric proportions of the array of astrometric CCDs compared to the slightly smaller spectroscopic array (spectroscopic CCDs are contained in 4 of the 7 rows of CCDs). For dimmer stars this number drops and for simplicity, we assume a constant 50\% chance of any scan having an associated spectroscopic measurement in our simulations (e.g. in Appendix \ref{ap:rvs_simulated_ast}).


We can define the spectroscopic reduced chi squared
\begin{equation}
\label{eq:ruwe_guess_conversion}
\widehat{\chi^2}_{spec}=\frac{1}{\nu}\sum_i\frac{\left(v_{r,i}-\langle v_r\rangle\right)^2 }{\sigma_{spec}^2}=\frac{RVN}{RVN-1}\frac{\sigma_{vr}^2}{\sigma_{spec}^2}
\end{equation}
where $v_{r,i}$ is the radial velocity measured on the $i^{th}$ observation and $\langle v_r \rangle$ is the mean of these values. $\nu=RVN-1$ is the number of degrees of freedom (our model being single constant value for the radial velocity of the star).

Finally we can convert this to the full analogue of $\textit{RUWE}_{ast}$:
\begin{equation}
\label{eqn:RUWE_Spec}
RUWE_{spec}=\sqrt{\widehat{\chi^2}_{spec}}=\sqrt{\frac{RVN}{RVN-1}}\frac{\sigma_{vr}}{\sigma_{spec}}
\end{equation}

If our model of a constant radial velocity is correct for the majority of sources the modal value of the reduced chi squared, and of $\textit{RUWE}_{spec}$ should be equal to one. We expect this to be the case, as even though binaries are ubiquitous only a subset of them will cause noticeable extra radial velocity variation.

Thus though we do not know $\sigma_{spec}$, we can use an initial guess (for example 1 km/s) and calculate a first guess $\textit{RUWE}_{spec}$ for each source. We can then estimate the real spectroscopic error as that which would set the modal value of the reduced chi squared to one. In this case we do this as a function of apparent magnitude ($m_G$) and colour ($m_{BP}-m_{RP}$) such that:
\begin{equation}
\begin{aligned}
\sigma_{spec}&(m_G,m_{BP}-m_{RP}) = \\
&mode\Big(RUWE_{spec}\big(m_G,m_{BP}-m_{RP}|\sigma_{spec}=1\ \rm{km/s}\big)\Big) \cdot 1\ \rm{km/s}.
\end{aligned}
\end{equation}

Calculating the mode is not entirely straightforward, so we replicate the method used in \citet{LindegrenRUWE}. If our sample were entirely composed of single stars with no other sources of noise we might expect the median value to correspond to the mode. However binaries will bias our distribution to higher values, thus we take the value at the 41$^{st}$ percentile to be the mode of the single star behaviour (i.e. assuming around 20\% of sources have been biased high by binarity or other sources of excess noise). 

Towards the edge of the parameter space (in $m_G$ and $m_{BP}-m_{RP}$) we have a diminishing number of data points and the question of how to estimate this error at, or beyond, the edge of the parameter space becomes difficult. To do this we use a novel method, an alternative to the analysis in \citet{LindegrenRUWE} which using splines to estimate the behaviour of the mode beyond the space spanned by the data. Our method is as follows:
\begin{itemize}
\item Firstly we define the span of our data ($2< m_G<18$ and $-2<m_{BP}-m_{RP}<8$) and renormalise these coordinates to a unit box spanning 0 to 1 in each dimension. Data points outside of this range are ignored.
\item We choose some subset of points, and for every point in the unit box we find which of this subset is its nearest neighbour.
\item For each of the subset of points we calculate the 41$^{st}$ percentile of the parameter of interest (our initial guess for $\textit{RUWE}_{spec}$) from all points for which it is the nearest neighbour.
\item This provides a Voronoi-like mesh of cells, with estimates for the parameter of interest spanning the entire unit box, with the inferred value at some point in the space being that calculated for the enclosing Voronoi cell.
\item We sample this on an equally spaced 100 by 100 grid (for easy interpolation later) spanning the unit box.
\item We repeat the above procedure with different random subsets, and the eventual values taken for the grid are the median across all runs.
\item Finally we can map our grid in the unit box back on to the original span of our data, to give a uniform grid of estimates of the local value of our parameter of interest evenly spanning the whole parameter space.
\end{itemize}

Note that though this process is closely related to the construction of a Voronoi tesselated cells, we do not have to actually construct this tesselation to perform the calculation. We perform all nearest neighbour calculations with \texttt{scipy.spatial.KDTree} which allows quick lookup of which of our roughly 6 million data points are closest to each of our subset of 1000 random points.

There are two (related) complications to the above procedure, the choice of the random subset of points and the number of repetitions. More repetitions will provide a smoother grid, where at the lower limit of no repetitions we get back exactly the Voronoi cell structure of any single iteration of the procedure. The only significant downside to more repetitions is computational cost, but here we find 10 to be a reasonable medium (running in minutes on a single CPU core).

We might be tempted to choose the subset of points randomly from the whole dataset - but this has a disadvantage: it is (linearly) weighted to select more points in areas of higher density. This is close to opposite to what we want, which is an estimate that is robust in areas of low density. Instead we adopt an extra step where we first estimate the density of sources and then weight our random selection by the density to some power, i.e.
\begin{equation}
\label{eqn:density_weight}
w_i=\frac{\rho_i^n}{\sum_i \rho_i^n}
\end{equation}
where $\rho_i$ is the density of sources in some local region around the $i^{th}$ point.

When $n$ is equal to zero all data points are equally likely to be selected and we are in the regime described above where sampling is proportional to density. Another way of putting this is that every cell can be expected to contain roughly the same number of data points.

At the other extreme, when $n=-1$, the chance of a point falling in some region being chosen is independent of the local density, and hence points are uniformly distributed in space (from the subset of the space spanned by the data). In some regions this will lead to cells which contain very few data points and (at least for small numbers of iterations) may be dominated by noise.

Larger (negative) values of $n$ could be chosen but these would give more samples to areas of lower density which seems counterproductive in most use cases.

We have settled upon $n=0.5$ as a reasonable compromise that well samples both dense and sparse regions.

This procedure can be seen in Figure \ref{fig:renormalised_spec_grid}, where we have made the conversion specified in Equation \ref{eq:ruwe_guess_conversion} to convert directly to the inferred spectroscopic error. We can see the cell-like structure for a single iteration, and the degree to which these cells span the parameter space. Reassuringly as the number of iterations increases the agreement between the different methods of selecting random points improves - suggesting that repeated samplings alone may be sufficient even if we had used the naive selection of points at random (without weights).

Thus we have a smooth estimate of our parameter of interest (here the spectroscopic error) which can easily be interpolated anywhere on this grid, and could even be extrapolated beyond.




\subsection{Photometric error}
\label{sec:renormalization_ruwe_phot}
We can repeat an analogous calculation to that which we've applied to spectroscopic errors to \textit{Gaia} photometry. This allows us to estimate if the flux of any given star is varying by more than we would expect for its colour and magnitude. In some cases this is a measure of unreliable data, whilst in others this variation can be astrophysical, associated with any of a number of sources of variability.

We start with the same catalogue of sources, and use the \texttt{PHOT\_G\_MEAN\_FLUX\_ERROR} and \texttt{PHOT\_G\_N\_OBS}, analogous to $RVE$ and $RVN$. There is no noise floor to subtract for photometric errors, but otherwise each step is the perfect analogue of those presented above.

Figure \ref{fig:sigmaphot_bprp_g} shows this process. Because we are working directly in the count of electron per second (the raw data read from each CCD) the error is largest for the brightest stars. While the integration time per observation is shorter for stars brighter than $m_G\approx 12$, with increasingly shorter integration times for brighter stars such that ideally the signal is the same for all stars, in practice we see that the Poisson error ($\sim \sqrt{N}$ for $N$ photons) is not flat with magnitude.
If we were to calculate the relative error (i.e. dividing through by $N$) we would expect the bright stars to be most accurately measured. The dependence on colour is slight, but there is some variation and an apparent change in behaviour above and below a $m_{BP}-m_{RP}$ of $\sim$1.



\section{Threshold values for $\textit{RUWE}_{spec}$ and $\textit{RUWE}_{phot}$}
\label{ap:threshold_ruwe}
\subsection{Distinguishing binaries from single sources using $\textit{RUWE}_{spec}$}
\label{ap:rvs_simulated_ast}
To estimate the value of $\textit{RUWE}_{spec}$ that can be used to distinguish single sources from binaries, we simulate another population of 100,000 MS-MS binaries and 100,000 single sources using $\code{astromet.py}$. With the exception of periods, masses, and light ratios, almost all parameters are randomly drawn from the distributions provided in Table \ref{table:simulated_parameter_distribution}; changed parameter distributions are provided in Table \ref{table:MS_simulated_parameter_distribution}. The mass distribution for main sequence stars is from \cite{Chabrier05}. We choose a log-uniform period distribution for binaries spanning a day to a 100 years--the regime we expect spectroscopic errors to be sensitive to--and the light ratio $l=q^{3.5}$ is estimated from the main-sequence luminosity relation. For simplicity, we assume a probability of 50\% of each set of astrometric measurements also yielding radial velocity measurements.  

\begin{figure}
\includegraphics[width=0.49\textwidth]{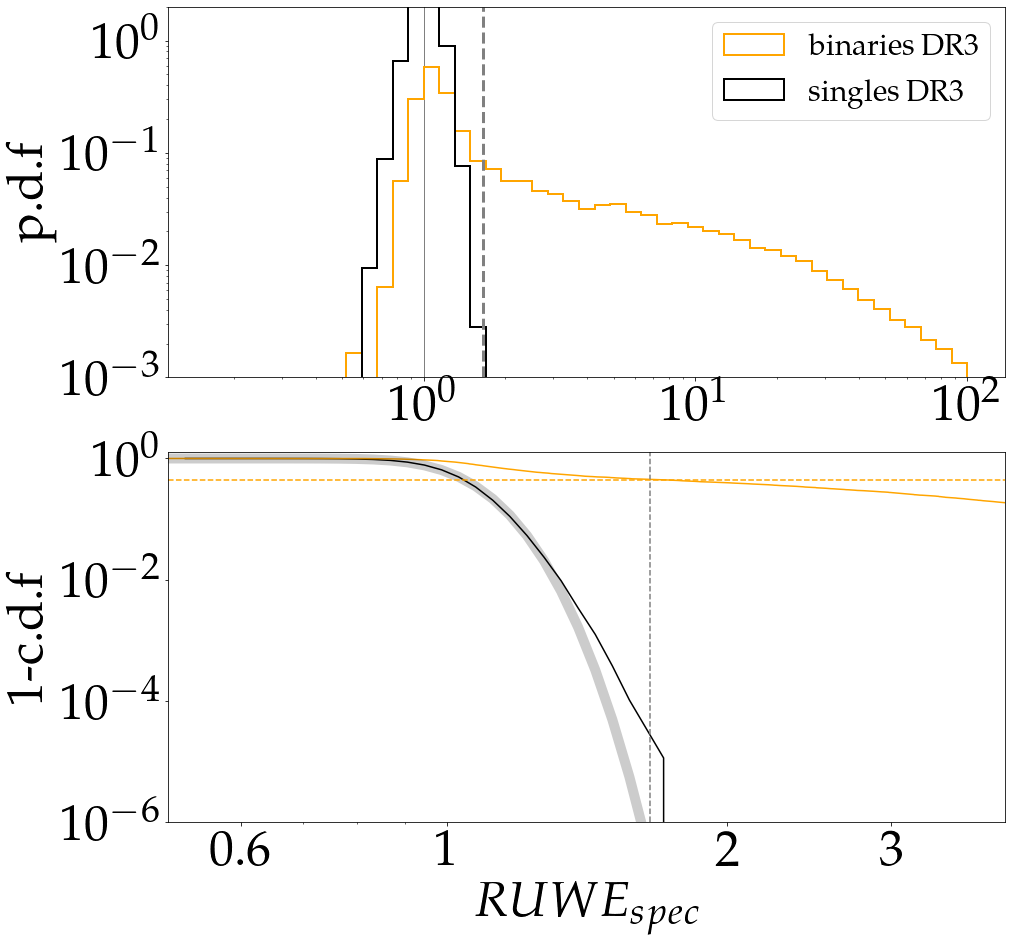}
\caption{Top: distribution of DR3 $\textit{RUWE}_{spec}$ (solid) for simulated single sources (grey) and binaries (yellow), with grey vertical lines at $\textit{RUWE}_{spec}$ = 1 and $\textit{RUWE}_{spec}$=1.7. Bottom: survival function for single sources (grey) and binaries (yellow). A $\chi$ distribution is fit to the single source curve and shown with the thick grey line. The grey vertical line is at 1.7, and the yellow horizontal line is plotted at 0.4.}
\label{fig:RUWE_RVS_simulated_dist}
\end{figure}

Figure \ref{fig:RUWE_RVS_simulated_dist} shows the distribution of $\textit{RUWE}_{spec}$ for simulated single systems and binaries over the DR3 time baseline. We see that the distribution of single sources peaks is significantly narrower than the distribution of binaries, and peaks at a $\textit{RUWE}_{spec}$ of 1. \cite{Penoyre20} showed that beyond $\textit{RUWE}_{ast}>1.25$--the cut recommended to separate single sources from binaries--approximately one in a million single sources remain. We conduct a analogous analysis by fitting a $\chi$ distribution to the $\textit{RUWE}_{spec}$ distribution of single sources, and find that only one in a million single sources have $\textit{RUWE}_{spec}>1.7$. Because these exact values are sensitive to the number of radial velocity measurements--which we have approximated to be 1/2 of total \textit{Gaia} observation times per source--we adopt a critical value of $\textit{RUWE}_{spec}=2$ for DR3.

\begin{deluxetable}{c c c}
\tabletypesize{\footnotesize}
\tablecolumns{8}
\tablewidth{\linewidth}
\tablecaption{MS-MS Parameter Distribution \label{table:MS_simulated_parameter_distribution}}
\tablehead{\colhead{Parameter} & \colhead{Description} & \colhead{Distribution}} 
\startdata
$P$ (days) & period & $10^{\mathcal{U}(-2.56,2)}$ \\
$m_1$ ($M_\odot$) & mass of primary &$10^{\mathcal{N}(-0.66,0.57)}$ \\
$m_2$ ($M_\odot$) & mass of secondary &$10^{\mathcal{N}(-0.66,0.57)}$ \\
$q$ & mass ratio  & $m_2/m_1$\\
$l$ & light ratio & $q^{3.5}$
\enddata
\vspace{-0.5cm}
\tablecomments{Distribution of changed parameters for the MS-MS binaries used in the analysis provided in Appendix \ref{ap:rvs_simulated_ast}. All other parameters are given in Table \ref{table:simulated_parameter_distribution}}. 
\end{deluxetable}


\subsection{Distinguishing variable from non-variable sources using $\textit{RUWE}_{phot}$}
\label{ap:ruwephot}
\begin{figure}
\includegraphics[width=0.49\textwidth]{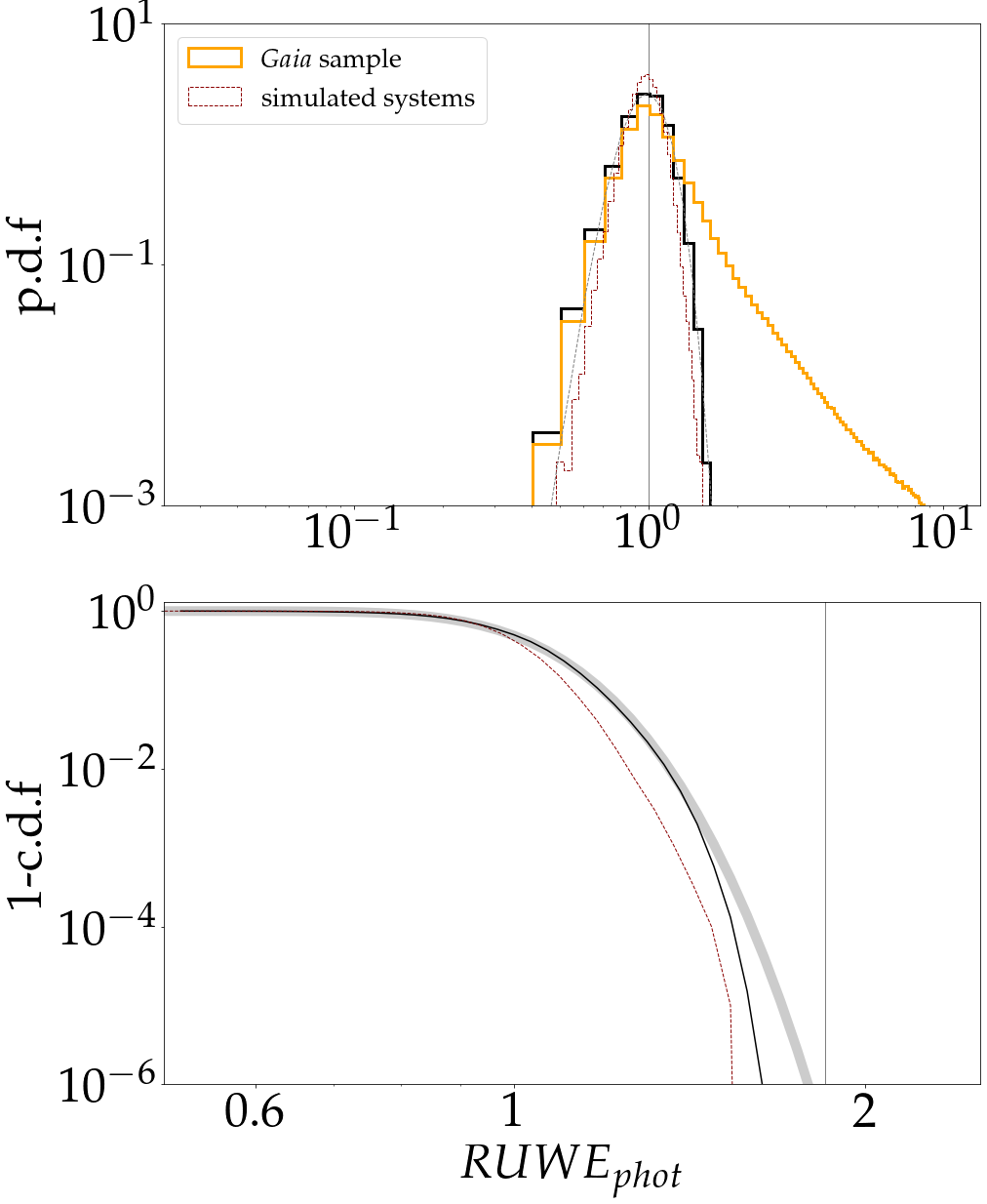}
\caption{Top: distribution of DR3 $\textit{RUWE}_{phot}$ for RVS sample (orange) and simulated non-variable sources (dark red), with a grey vertical line at 1. The solid black curve shows the modeled distribution of non-variable sources, obtained by reflecting  sources in the RVS sample with $\textit{RUWE}_{phot}<1$ about the peak (at $\textit{RUWE}_{phot}=1$). Bottom: survival functions with a $\chi$ distribution fitted to the modeled distribution of non-variable sources shown in grey. A vertical line is plotted at $\textit{RUWE}_{phot}=1.9$.}
\label{fig:RUWE_phot_survival_func}
\end{figure}

To estimate the value of $\textit{RUWE}_{phot}$ that can be used to distinguish non-variable from variable sources, we plot the distribution of $\textit{RUWE}_{phot}$ (calculated in Section \ref{sec:renormalization_ruwe_phot}) for our $RVS$ sample, and model the distribution of non-variable sources by reflecting sources with $\textit{RUWE}_{phot}<1$ about the the peak of the distribution ($\textit{RUWE}_{phot}=1$). Figure \ref{fig:RUWE_phot_survival_func} shows that there is good agreement between this modelled distribution and the distribution of simulated, (non-variable) systems with the same parameter distribution of single sources described in \ref{ap:rvs_simulated_ast} and a constant photometric error of $\sigma_{phot}=10^4$ es$^{-1}$.

We fit a $\chi$ distribution to the $\textit{RUWE}_{phot}$ distribution of modelled non-variable sources in the $RVS$ sample, and find that only one in a million single sources have $\textit{RUWE}_{phot}>1.9$. We adopt a critical value of $\textit{RUWE}_{phot}=2$ for distinguishing non-variable from variable sources in DR3.

\label{lastpage}

\end{document}